# Applications of exchange coupled bi-magnetic hard/soft and soft/hard magnetic core/shell nanoparticles

A. López-Ortega[1], M. Estrader[2], G. Salazar-Alvarez[3], A.G. Roca[4], J. Nogués[4,5]

[1]INSTM and Dipartimento di Chimica "U. Schiff", Università degli Studi di Firenze, Via della Lastruccia 3, Sesto Fiorentino, I-50019, Firenze, Italy.
[2]Departament de Química Inorgànica, Universitat de Barcelona, Diagonal 645, E-08028, Barcelona, Spain.
[3]Department of Materials and Environmental Chemistry, Arrhenius Laboratory, Stockholm University, S-10691 Stockholm, Sweden.
[4]ICN2 - Institut Catala de Nanociencia i Nanotecnologia, Campus UAB, E-08193 Bellaterra (Barcelona), Spain.
[5]ICREA - Institució Catalana de Recerca i Estudis Avançats, Barcelona, Spain.

## ABSTRACT

The applications of exchange coupled bi-magnetic hard/soft and soft/hard ferromagnetic core/shell nanoparticles are reviewed. After a brief description of the main synthesis approaches and the core/shell structural-morphological characterization, the basic static and dynamic magnetic properties are presented. Five different types of prospective applications, based on diverse patents and research articles, are described: permanent magnets, recording media, microwave absorption, biomedical applications and other applications. Both the advantages of the core/shell morphology and some of the remaining challenges are discussed.

## Contents

∗ Corresponding author at: ICN2 - Institut Catala de Nanociencia i Nanotecnologia, Campus UAB, E-08193 Bellaterra (Barcelona), Spain.
E-mail address: Josep.Nogues@uab.cat (J. Nogués).

## 1. Introduction

Magnetic nanoparticles are gaining increasing interest equally in industry and research due to both the numerous applications in very widespread fields, ranging from engineering (e.g., magnetic recording media or magnetic seals) to biomedical applications (e.g., magnetic resonance imaging, drug delivery or hyperthermia) and their appealing novel properties [1–16]. Interestingly, the advances in synthetic chemistry allowing the extraordinary control of the growth parameters has led to the development of more advanced magnetic nanoparticles comprising two (or more) materials such as core/shell particles [17–32]. These types of multiphase nanostructures can combine the different functionalities (*e.g.*, catalytical, optical, magnetic or biomedical) of the diverse constituents bringing about novel and enhanced properties which are resulting in innovative applications of magnetic nanoparticles. A particularly interesting topic in core/shell magnetic nanoparticles is the study of bi-magnetic core/shell nanoparticles, i.e., where both the core and the shell exhibit magnetic properties (ferromagnetic (FM), ferrimagnetic (FiM) or antiferromagnetic (AFM)). In these systems the exchange interaction between both constituents brings about an extra degree of freedom to tailor the overall properties of the nanoparticles. Since the discovery of exchange bias (i.e., the loop shift in the field axis of the hysteresis loops [33,34]) in Co/CoO nanoparticles [35], FM/AFM and inverse AFM/FM core/shell nanoparticles have been extensively studied [36–40] Interestingly, less attention has been paid to FM or FiM "conventional" hard/soft and "inverted" soft/hard, core/shell nanoparticles (see Fig. 1) although it has been demonstrated for bulk and thin film systems that these bi-component materials can exhibit very appealing properties [41–46]. However, in recent years substantial advancement has occurred in this field, particularly in permanent magnets [47–71], magnetic recording media [64,72–79], microwave absorption [80–82], ferrofluids [83] or biomedical applications [84–88], where it has been shown that for certain applications the use of bi-magnetic core/shell nanoparticles can be advantageous over single magnetic nanoparticles. Here we present a review of the current state of the research in the topic of hard-soft core/shell nanoparticles. After introducing the main synthesis approaches, we discuss some of the basic phenomenology of hard-soft core/shell nanoparticles. Later we overview some of their main potential applications (in permanent magnets, magnetic recording, microwave absorption or biomedicine), finishing with some conclusions.

## 2. Basic Phenomenology

*2.1 Synthesis of Hard-Soft Nanoparticles*

There exists a great variety of fabrication methods to obtain inorganic core/shell nanoparticles based on chemical and physical approaches [17–30] (see Table 1). However, given the great versatility of many chemical routes in controlling materials, crystallinity, homogeneity, sizes or even shapes, less attention has been paid to the synthesis of core/shell nanoparticles using physical approaches [17–30]. Some examples of standard chemical routes used for the synthesis of hard-soft core/shell nanoparticles are co-precipitation [89–91], thermal decomposition [92–109], metal reduction [110,111], microwave-assisted methods [112,113] and electrodeposition [114,115].

One of the simplest routes to form inorganic core/shell nanoparticles arises from the surface treatment of the nanoparticles [89,116–130]. This approach has been probably the most developed method to form core/shell nanoparticles and specifically on the study of ferromagnetic/antiferromagnetic systems [36]. This procedure consists on the surface treatment (e.g., oxidation, reduction, nitration or carbidization) of the nanoparticles, leading to the formation of a layer on their surface with a different phase (i.e., with dissimilar physiochemical properties of those of the core). In the case of transition metal nanoparticles, controlled oxidation leads to the formation of an oxide passivation layer. Similarly, for oxide nanoparticles oxygen passivation can give rise to the formation of an oxide shell with a higher oxidation state. On the other hand, controlled reduction of an oxide nanoparticle may lead to a metallic surface layer. Despite its simplicity, for the specific case of hard-soft core/shell nanoparticles this approach has not been extensively exploited. This stems from three main drawbacks of this method, (i) the shell is always derived from the core, consequently their sizes cannot be independently controlled; (ii) since the shell phase has to be derived from a surface treatment of the core material, the possible choices of core/shell phases are somewhat limited; (iii) since the materials in the core and the shell often have quite different structural characteristics (e.g., metal vs. oxide), the structural quality of the shell may be inferior. Despite these disadvantages, several examples of this approach using surface oxidation, mainly for inverse soft/hard systems, can be found in the literature: FeCo/$CoFe_2O_4$ [120,121], FeCoB/$CoFe_2O_4$ [123] and FeCo(Al)/FeCo(Al)$O_x$ [124]. Concerning controlled reduction, only the $CoFe_2O_4$/FeCo system has been studied, although two different reduction approaches have been used: hydrogen reduction [125–129] and carbon reduction [130]. Interestingly, previous to surface treatment the nanoparticles were synthesized mainly by wet chemistry methods such as: thermal decomposition [122], metal reduction [123], sol-gel [129] or co-precipitation [126]. However, some physical approaches have been also reported to obtain the initial nanoparticles such as: inert gas condensation [120], thermal plasma method [124], flame spray pyrolysis [121] and ionic coordination [125–128]. Notably, some synthesis approaches may lead to the desired core/shell structure without the need of post-synthesis treatments due to the non-homogeneous growth conditions [131,132].

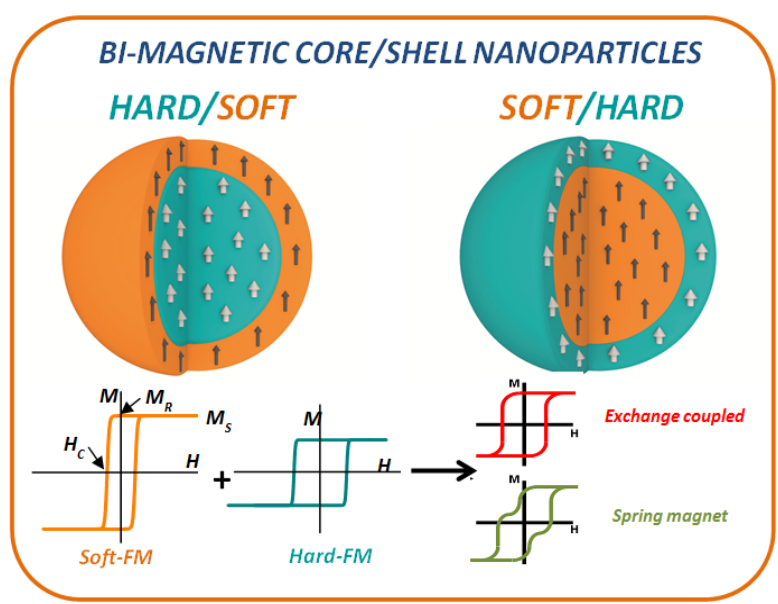

**Fig. 1.** Schematic representation of the hard/soft and soft/hard core/shell structures (top). Simple graphic representation of the hysteresis loops for soft and hard FM and their possible combinations: exchange coupled and exchange-spring magnet. The saturation magnetization, $M_S$, the remanent magnetization, $M_R$ and the coercivity $H_C$, are indicated in the soft-FM loop.

**Table 1**
Summary of the pros and cons of the main chemistry approaches to synthesize inorganic bi-magnetic core/shell nanoparticles.

| Synthetic method | Pros | Cons |
| --- | --- | --- |
| Based on seeds | | |
| Surface treatment | Easy, inexpensive | Lack of possible combinations and bad crystallinity of shell structure and interphase |
| Seeded-growth: Thermal decomposition | Great versatility of possible combinations | Difficult, expensive, organic solvents |
| Seeded-growth: Sol-gel, microemulsion and co-precipitation | Inexpensive, versatility of combinations, aqueous media | Difficult shell growth control |
| Ion substitution | Easy, aqueous media | Limited available shell structures |
| Other | | |
| Self-assembly | High degree of crystallinity and good interphase | Difficult, expensive and lack of possible combinations |
| Microwave irradiation | Easy, inexpensive | Restricted number of possible combinations |

Another interesting approach is the seeded-growth method. This method is particularly appealing since it allows synthesizing heterostructures (i.e., with core and shell composed of different materials) with an exquisite control of the sizes and morphologies. This method is based in a two-step synthesis process, where pre-made nanoparticles are used as seeds for the posterior deposition of the shell. From a thermodynamic point of view, this approach is controlled by the heterogeneous nucleation of the shell precursor avoiding, thus, the homogeneous nucleation of new nanoparticles [105,107,133,134]. In comparison with the surface treatment approach this method allows an independent control of the core (pre-made nanoparticles) and the shell during the synthesis, resulting in a controllable size ratio between the core diameter and shell thickness and allowing the growth of shells with excellent crystallinity. Specifically, in the case of core/shell structures based on systems formed by hard-soft magnetic phases the main efforts have been focused in the synthesis of nanoparticles based on $FePt/MFe_2O_4$ (where M = Fe, Co and Zn) core/shell systems [92–99,135–138]. Interestingly, many of these systems are initially composed of two soft magnetic phases (i.e., atomically disordered fcc-FePt and iron based spinel structures) and posterior heat treatments permit the formation of the ordered $L1_0$-FePt hard ferromagnetic phase, thus leading to the desired hard/soft structure. Usually the synthesis of the fcc-FePt seeds is performed through the thermal decomposition of $Pt(acetylacetonate)_2$ and either $Fe(CO)_5$ [92–99,135–138] or $Fe(acetylacetonate)_2$ precursors [94,95,98]. The shell growth has been mainly carried out through the two-steps seeded-growth approach by the thermal decomposition of iron and other transition metal salts [92–99]. Interestingly, the use of excess of iron precursor, $Fe(CO)_5$, during the synthesis of FePt nanoparticles can also promote the growth of an iron oxide shell in the so-called one-pot synthesis [135–138]. Remarkably, originally, the purpose of the growth of spinel oxide shells was preventing the FePt nanoparticles coalescence during the annealing. However, magnetic characterization demonstrated the exchange coupling between both types of FePt phases (i.e., fcc pre- and $L1_0$ post-annealed) and the oxide shell [92–95].

Although the study of $FePt/MFe_2O_4$ (M= Fe, Co and Zn) core/shell nanoparticles has caught a huge attention, other different hard-soft core/shell systems have also been obtained by the seeded-growth approach. Some examples synthesized by the thermal decomposition of the shell precursor are $Fe/CoFe_2O_4$ [100], $Fe/Fe_3O_4/FePt$ [139], $CoFe_2O_4/ZnFe_2O_4$ [101], $CoFe_2O_4/MnFe_2O_4$ [102,103], $ZnFe_2O_4/CoFe_2O_4$ [101], $MnFe_2O_4/CoFe_2O_4$ [102] , $Sm(Co_{1−x}Fe_x)_5/Fe_3O_4$ [104], $Mn_xFe_{3−x}O_4/Fe_xMn_{3−x}O_4$ [105], $Fe_3O_4/Mn_3O_4$ [106–108] and $Mn_3O_4/Fe_3O_4$ [107].
Notably, although the thermal decomposition method is probably the most common approach to grow shells, other different wet chemistry routes have also demonstrated their suitability to perform this purpose. Some examples are co-precipitation ($CoFe_2O_4/Fe_3O_4$ [90] and $SrFe_{12}O_{19}/CoFe_2O_4$ [91]), reduction ($Nd_2Fe_{14}B/FeCo$ [110] and $Nd_2Fe_{14}B/\alpha$-Fe [111]), electrodeposition ($SmCo_5/FeNi$ [114], $SmCo_5/FeCo$ [114], SmCo/Co [115] and SmCo/FeCo [115]) and polymerization ($CoFe_2O_4/Ni_{0.5}Zn_{0.5}Fe_2O_4$ [140]).

It should be pointed out that other less common approaches to synthesize inverse and conventional hard-soft core/shell structures have also been reported. For example, the previous mentioned one-pot synthesis of $FePt/Fe_3O_4$ core/shell nanoparticles where the use of an excess of iron precursor in the synthesis of FePt nanoparticles leads the formation of an iron oxide shell [135–138]. A rather unique approach is the use of $Fe_3O_4$ nanoparticles self-assembled together with either Co or FePt nanoparticles. The annealing of these bi-modal self-assembled structures demonstrated the possibility to create $Fe_3O_4/CoFe_2O_4$ core/shell [141] and $Fe_3Pt/FePt$ nanocomposites [142], respectively. These systems are formed through ion substitution, i.e., Co and Pt ions substitute Fe ions at the surface of the $Fe_3O_4$ nanoparticles, respectively. Other approaches based on ion exchange or ion absorption use $Fe_3O_4$ nanoparticles dispersed in a medium containing Co ions, which can be superficially either absorbed or exchanged, thus creating a $CoFe_2O_4$ shell [143,144].

Thus far, the chemical methods used for the synthesis of hard-soft core/shell nanoparticles have been reviewed. Nevertheless there are many other chemical routes which, although they have not been used for hard-soft materials, have demonstrated their suitability to synthesize different types of inorganic/inorganic core/shell nanoparticles [17–31]. For instance, microemulsion and sol-gel approaches are probably two of the most useful routes to synthesize this type of architectures [145–152]. Moreover, other more unusual methods (sometimes specific for a given system) can also be found in the literature such as one-pot sonochemical synthesis of $Fe_3O_4/FeP$ core/shell nanoparticles [153] or microwave irradiation synthesis of Ni/Cu core/shell nanoparticles from initial mixed solution containing both precursors [112].

Finally, it should be emphasized that although less attention has been paid to the use of physical routes to synthesize inorganic/inorganic core/shell structures some examples can indeed be found in the literature, such as: radiolysis [154], simultaneous inert gas condensation and laser ablation [155–157], reactive gas condensation [158,159] pulsed laser ablation [160,161], sputtering [162–164], pyrolysis [165–167], thermal plasma [168,169] and ball milling [170,171].

*2.2 Structural-Morphological Characterizations*

Given the critical importance of the structure and the morphology (e.g., particle size distribution, core diameter, shell thickness or sharpness of the interphase) on the magnetic properties of core/shell hard-soft nanoparticles, a precise characterization is often required to understand them. Although there are no generally applicable methods to determine the structure of (core/shell) nanoparticles [172], a combination of traditional characterization approaches such as X-ray diffraction, neutron diffraction and transmission electron microscopy (TEM) with other sophisticated techniques are often needed to fully elucidate the structure and composition of bi-magnetic core/shell nanoparticles. In this section we review some of these techniques.

*2.2.1 Structure*
The structural determination of the core/shell nanoparticles is mostly carried out using diffraction methods, based on X-ray and neutron powder diffraction, or selected-area electron diffraction. A development in diffraction approaches depends on the analysis of the total scattering data ($0 \leq Q \leq 25$ Å$^{-1}$), which involves the extraction of the structure factor and the radial distribution function or, rather, the pair-distribution function in a given material. This type of analysis provides information on the local structure, making it a powerful tool for the study of heterogeneous nanoparticles, amorphous systems and other crystallographically-challenging materials [173]. Frison et al. used total scattering analysis to investigate the structure and relative composition of $Fe_3O_4/\gamma-Fe_2O_3$ core-shell nanoparticles, a very complicated system to study with conventional techniques [174].

Alternatively, X-ray absorption spectroscopy, in the extended X-ray absorption fine structure (EXAFS) modality (available only at synchrotron facilities), provides the radial distribution function around a central (X-ray absorbing) atom. Similarly to pair-distribution function, EXAFS can give information on the local structure around a given element. For instance, Baker et al. used EXAFS to study the degree of alloying in palladium-embedded iron nanoparticles, which results in Fe/Fe$_x$Pd$_{1-x}$ core/shell nanoparticles [175], whereas Gomes et al. used it to investigate the cation distribution in ZnFe$_2$O$_4$/γ-Fe$_2$O$_3$ core/shell nanoparticles [176].

*2.2.2. Morphology*

It is worth mentioning that although powder diffraction methods are relatively well-known, recent developments have focused towards the analysis of anisotropic line broadenings arising from shape anisotropy [177,178] as well as from defects or compositional variation [179]. Hence, a great deal about the defect structure and particle morphology can be inferred from conventional X-ray diffractograms simply by determining the crystal domain size along different crystallographic directions. Although the analysis is not trivial for very small particles due to peak broadening, in the case of core/shell nanoparticles from the volume ratio of the components it is possible to determine indirectly the morphology of the crystallites if a secondary technique is available, e.g., TEM[38,90].

However, particle morphology has been typically assessed from electron micrographs, where TEM is most useful to locally measure particle sizes and determine morphologies of nanoparticles [180]. With regard to core/shell nanoparticles, due to the increasing intensity of the electron scattering with increasing atomic number, Z [181], conventional bright-field TEM images may be useful in distinguishing the core from the shell only if the difference in average atomic composition is relatively large, as is the case for metal/metal oxide compositions or metals with at least one row of separation in the periodic table, as shown for instance in refs.[135,175,182–186].

In the case of core/shell nanoparticles with weak contrast, dark-field (DF) TEM imaging can be used to differentiate the two components on the basis of structural dissimilarities. Moreover, scanning TEM (STEM) using the so-called high angular annular dark field mode (HAADF) can enhance the elemental contrast as the image intensity, I, scales with near the square of the atomic number, i.e., $I \propto Z^{1.7}$ [181]. For instance, Hai et al. [187] and Wetterskog et al. [188] visualized the core and the shell of Fe$_{1-x}$O/Fe$_{3-\delta}$O$_4$ nanocubes by comparing DF-TEM images taken using diffracted beams corresponding only to the spinel structure (Fe$_{3-\delta}$O$_4$) or both the spinel and rock salt structure (Fe$_{1-x}$O) (Fig. 2a,b) whereas Pichon et al. [189] and Liu et al. [190] used STEM-HAADF imaging to highlight the core from the shell (see Fig.2c-d). Lastly, it is also possible to acquire high-resolution TEM (HRTEM) images, perform a fast Fourier transform (FFT), and obtain the inverse FFT image by selecting specific diffraction spots, i.e., a kind of computer-processed DF-TEM (see Fig. 2g-i) [191]. A further development of this technique allows also the quantitative evaluation of the internal strains in such core/shell nanoparticles. This, so-called, geometric phase analysis of HRTEM images can result in 2D maps of the inner structure of for instance core/shell nanoparticles. Wetterskog et al. showed the presence of internal defects (antiphase boundaries) in Fe$_{1-x}$O/Fe$_{3-\delta}$O$_4$ nanocubes and how these internal defects could be directly correlated to the interface between the core and the shell (see Fig. 2e,f) [188].

Alternatively, spectroscopic TEM-based methods can also be employed to study core/shell nanoparticles. As a result of inelastic electron-matter interaction, the transmitted electrons may suffer a quantized energy loss which can be measured as an electron energy-loss spectrum (EELS) [192]. Transmission electron microscopes equipped with an EEL spectrometer can then be used to acquire element-specific images through energy filtering of the EEL spectra [181]. For instance Salazar-Alvarez et al. used EFTEM to image Ca. 1 nm thick MnO shell deposited on FeO/Fe$_3$O$_4$ cores (see Fig.3a-d) [193] and Ong et al. employed it for the characterization of Fe/FeO/Fe$_3$O$_4$ nanoparticles [182]. An improvement to this technique is the combination of STEM with a very small probe size, typically <1 nm and EELS, which provides with lateral resolution at the atomic level [194,195]. This technique has been utilized to image various types of core/shell nanoparticles [107,196,197] (see Fig. 3e). Importantly, in combination with HRTEM images, this approach allows visualizing the sharpness of the interface [107].

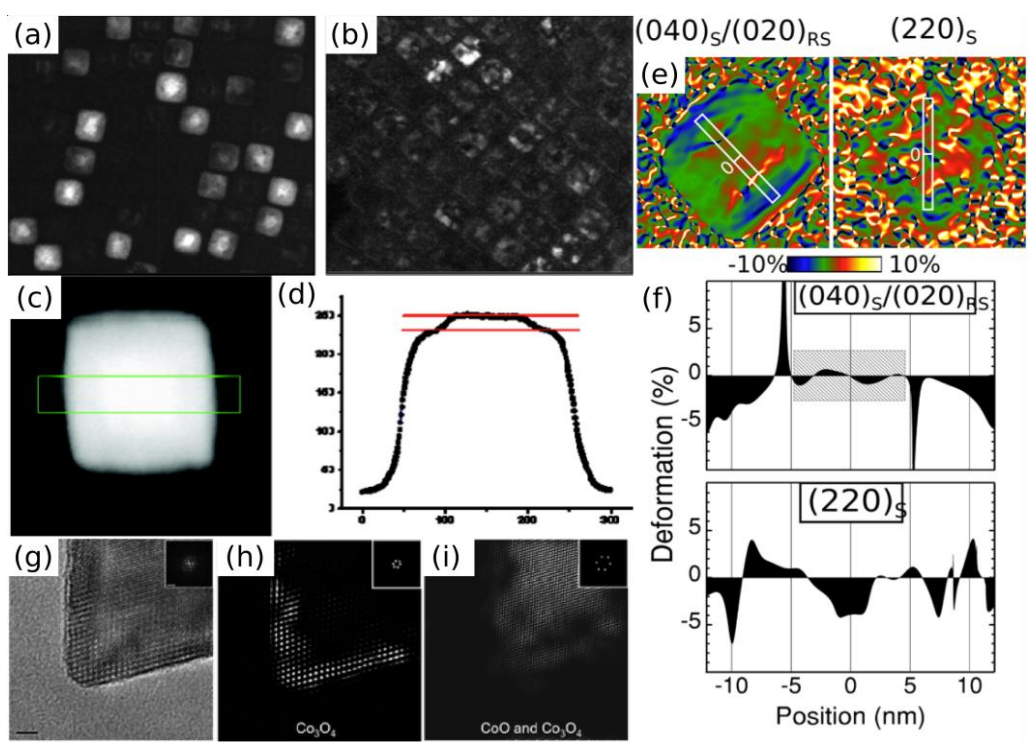

**Fig. 2**. Dark field-TEM images of $Fe_{1-x}O/Fe_{3-\delta}O_4$ core/shell nanoparticles. (a) Image taken using the $(400)_{Spinel}/(200)_{Rock\ salt}$ diffraction beam, (b) image taken using the $(220)_{Spinel}$ diffracted beam [187]. (c) STEM-HAADF image and the corresponding averaged line profile (d) of the green area; the red lines highlight the difference between the core and shell areas [189]. (e) Lattice deformation maps of a $Fe_{1-x}O/Fe_{3-\delta}O_4$ core/shell nanocube obtained by geometric phase analysis of the $(040)_{Spinel}/(020)_{Rock\ salt}$ and the $(220)_{Spinel}$ reflections oriented parallel to the white rectangle. (f) Lattice deformation profiles obtained by integration from the bottom to the top of the white rectangles in (e) [188]. (g) HRTEM image (scale bar = 2 nm) of the tip of a $CoO/Co_3O_4$ octahedron. (h) and (g) FFT-filtering of the image in (i) using diffraction spots that correspond only to $Co_3O_4$ and CoO, respectively [191]. (For interpretation of the references to color in this figure legend, the reader is referred to the web version of this article.)
*Source*: Panels (a) and (b) were reprinted with permission from Ref. [187].
© 2010, by Elsevier.
Panels (c) and (d) were reprinted with permission from Ref. [189].
© 2013, by the American Chemical Society.
Panels (e) and (f) were reprinted with permission from Ref. [188].
© 2011, by the American Chemical Society.
Panels (g), (h) and (i) were reprinted with permission from Ref. [191].
**© 2014, by the American Chemical Society.**

TEM based techniques have the drawback that usually only a handful of particles are analyzed. To circumvent this problem several other techniques have been used for the reconstruction of the core/shell configuration of macroscopic amounts of sample to obtain averaged information. Here small-angle X-ray and neutron scattering (SAXS and SANS, respectively) have been useful to determine the particle size and particle morphology through the radius of gyration of the particle in a solvent [198]. Differentiation of core/shell configurations is possible if the average composition of the core and the shell are sufficiently large. Similar to electrons, the scattering of X-rays increases with the atomic number and thus it may be difficult to separate the contributions of the core and the shell if the average atomic compositions have a similar average atomic number. Two alternative options in such situations are SANS and anomalous

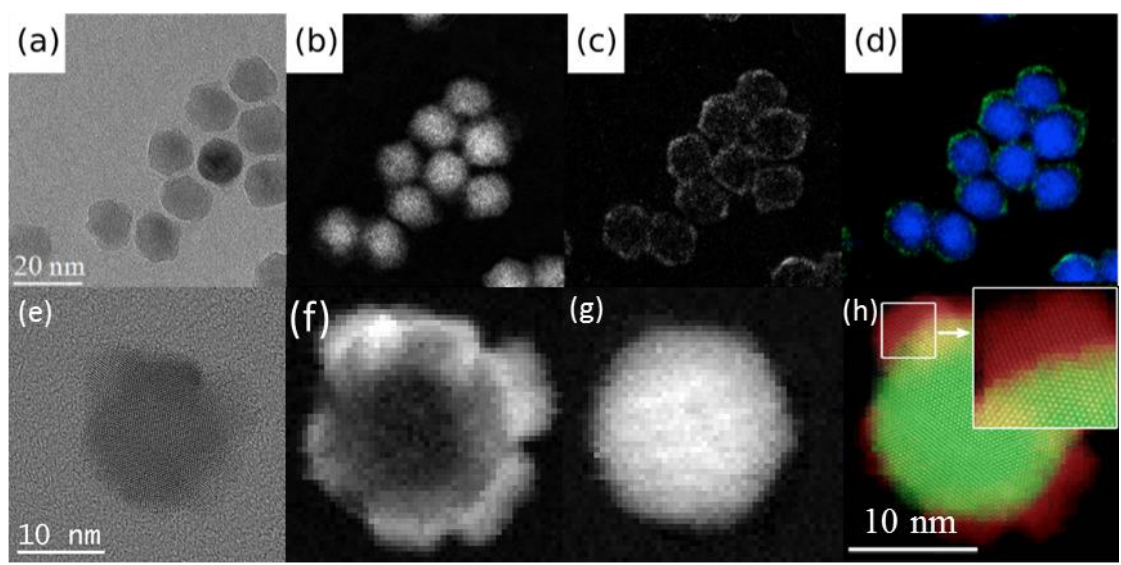

**Fig. 3** EELS-based TEM images. (a) BF-TEM image of FeO/Fe$_3$O$_4$/MnO onion nanoparticle, (b) and (c) EFTEM images of (a) acquired at the (b) iron and (c) manganese L$_3$ edges. (d) False color overlay image of (b) and (c) [193]. (e) High resolution-STEM-HAADF image of a Fe$_3$O$_4$/Mn$_3$O$_4$ core/shell nanoparticle, (f) and (g) STEM-EELS spectrum images at the Mn- and Fe-edges. (h) Overlaid image of the HAADF and spectrum images in false color. Shown in the inset is an enlarged view of the highlighted area [107]. (For interpretation of the references to color in this figure legend, the reader is referred to the web version of this article.)
*Source:* Panels (a)-(d) were reprinted with permission from Ref. [193].
© 2011, by the American Chemical Society.
Panels (e)-(g) were provided by Prof. M. Varela (ORNL-Univ. Complutense).
Panel (h) was reprinted with permission from Ref. [107].
© 2013, by Nature.

SAXS (ASAXS). In the case of SANS, the scattering intensity depends on the interaction of neutrons with the nucleus which is isotope- and spin-dependent and can also be positive or negative whereby scattering contrast between different components may be readily observed even for neighboring elements [198]. Moreover, as the neutrons also have a spin moment, they can be used to probe the magnetic scattering by the nanoparticles. It is then possible to distinguish magnetic differences between a core and a shell, both in the case of chemically distinct core/shell compositions [199] or in the case of magnetic core/shell architectures (i.e., where the core and the shell have the same composition but different magnetic properties) [200–202].

Regarding ASAXS, this technique exploits the increment of X-ray scattering intensity at the X-ray absorption edge of a given element. In this manner, by tuning the energy of the incident beam, it is possible to enhance the elemental contrast between neighboring elements. Krycka et al. demonstrated that not only it is possible to differentiate between the core and the shell in the Fe$_3$O$_4$/γ-Mn$_2$O$_3$ system but also that the scattered intensity can be modeled to estimate the degree of cation intermixing at the core/shell interface (see Fig. 4) [197].

### 2.2.3. Composition

Many bulk spectroscopic methods have been used to determine the accurate global composition of materials, such atomic absorption or emission [203]. However, their use tends to be destructive as the material is usually dissolved or digested. Instead other non-invasive bulk methods such as traditional X-ray fluorescence [203] and energy-dispersive X-ray spectroscopy (EDS) [181] are preferred albeit their lower spatial resolution.

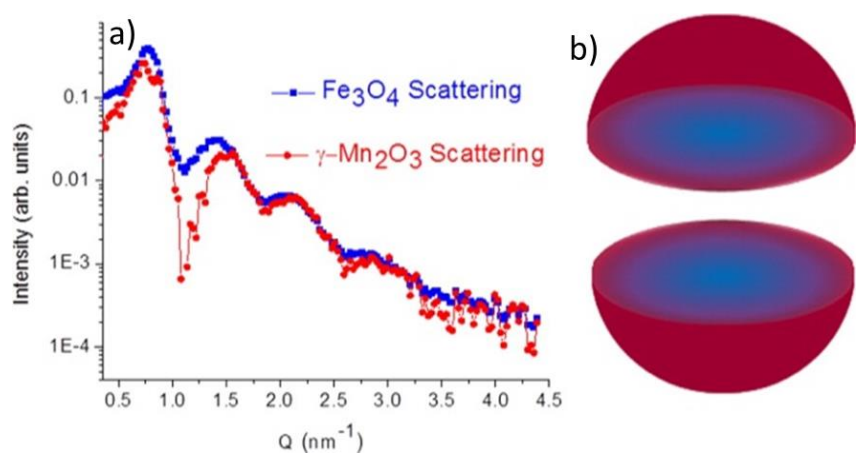

**Fig. 4**. (a) Material-specific ASAXS profiles for $Fe_3O_4$ and $\gamma$-$Mn_2O_3$ of $Fe_3O_4$/$\gamma$-$Mn_2O_3$ core/shell nanoparticles. (b) Cartoon depicting a core/shell nanoparticle with graded composition [197].
*Source*: Figure reprinted with permission from Ref. [197].
© 2013, by the American Chemical Society.

The local composition around a central atom can be obtained also with somewhat more specialized spectroscopic techniques such as X-ray absorption spectroscopy (XAS) (synchrotron-based technique) and EELS (TEM-based technique). Similar to EXAFS, it is possible to determine the composition and even the oxidation state of the elements from the energy shift. Nevertheless, whereas XAS has been routinely used for the determination of the composition and local environment of a given atom, EELS can also be used to probe the spatial elemental distribution and oxidation state in a given material, usually in combination with STEM (see Fig. 3) in order to achieve atomic resolution [204]. Moreover, recent developments in EDS detectors have resulted in STEM-EDS providing also atomic resolution similar to that of STEM-EELS [205].

Finally, Rietveld analysis of X-ray diffractograms can also be used to qualitatively probe the composition of core/shell nanoparticles. For instance, Salazar-Alvarez et al. [143] studied in $\gamma$-$Fe_2O_3$ nanoparticles the relative changes in the intensity of diffraction peaks corresponding to (surface) octahedral vacancies during their substitution with Co(II) ions. The surface increase was correlated with the changes in blocking temperature and corroborated with X-ray photoelectron spectroscopy.

A technique derived from XAS, where circularly polarized light is used (left-handed or right-handed) to probe the magnetization of nanoparticles is X-ray magnetic circular dichroism (XMCD – see also Section 2.3.1 Static Magnetic Properties). A typical XMCD pattern is obtained by subtracting the XAS absorption spectra with a given handedness from the other. The so-called XMCD sum rules are then applied to obtain quantitative information about the absolute magnetization of the material and its decomposition into orbital and spin moment [206,207]. Due to the characteristics of XAS, the technique is mostly a surface-probe and is ideal to probe nanoparticles [208], and especially suited to investigate core/shell [209] or binary nanoparticles [210]. Fauth [209] investigated the composition-dependent magnetic behavior of Fe/$\gamma$-$Fe_2O_3$ and FePt/$Fe_3O_4$ core/shell nanoparticles with a combination of experimentally-acquired and numerically-modeled XMCD spectra. Alternatively, Nolle et al. studied the composition-dependent magnetization of FePt/$FeO_x$ binary nanoparticles through a linear deconvolution of the XMCD spectra using reference data for FePt, $\gamma$-$Fe_2O_3$ and $Fe_3O_4$ nanoparticles [210]. Skoropata et al. used XMCD to study also the composition-dependent magnetization of $\gamma$-$Fe_2O_3$ nanoparticles with a nominal shell of Cu, CoO, MnO, and NiO, where it was found that there is a significant substitution of surface vacancies with the divalent metal ions [211].

*2.2.4. Rising techniques*

Lastly, amongst the rising techniques that have been reported very recently for core/shell magnetic materials are the resonant inelastic X-ray spectroscopy (RIXS)[212,213], and the atom-probe microscopy or tomography [214].

The RIXS processes are closely related to EXAFS although the energies used for the excitation are chosen so as to increase the inelastic scattering at a resonant frequency. During a RIXS process a core electron is excited to the valence band, the core hole is then filled by a valence electron and subsequently followed by the emission of a photon. This implies that in the absence of a core hole in the final state the energy of the photon out can be directly correlated with the valence band processes [212]. Moreover, the RIXS processes are polarization dependent, which result in an interesting technique to probe magnetization processes [213]. The advantages of RIXS (combining the chemical selectivity of core spectroscopies, the bulk sensitivity of hard X-rays, and the possibility to measure a large amount of particles) have been recently exploited to study the presence of interdiffused interfaces in $Fe_3O_4/Mn_3O_4$ core/shell nanoparticles [106].

Atom-probe microscopy or tomography consists of the field-driven evaporation of surface ions and their identification with a mass spectrometer. Knowing the geometry of the material, its distance to the ion detector, and the time of travel of the ions, it is possible to reconstruct in 3D the morphology of the original material at atomic resolution [214]. Atom-probe microscopy-tomography has been used to study, for instance, the compositional gradients in $SmCo_5$/α-Fe milled powders showing the formation of α-Fe nanoparticles in a $SmCo_5$ matrix, Cu-Co granular alloys and CuNi nanostructures [215,216].

Finally, another technique with great potential is EELS-Tomography. By acquiring EELS spectrum images (SI) in a TEM at different tilt angles (i.e., tomography), it is possible to obtain a 3D reconstruction of the elemental distribution in the material. For instance, EELS-SI tomography has been used to chemically determine the 3D structure of $Co_3O_4/Fe_xCo_{(3-x)}O_4$ core/shell mesoporous particles down to pore resolution [217].

*2.3 Basic Static and Dynamic Magnetic Properties*

*2.3.1 Static Magnetic Properties*

The main properties expected in hard-soft core/shell nanoparticles are closely related to the ones observed in hard-soft bilayers [41]. The behavior of these systems can be understood as a combination of the intrinsic parameters of the hard and soft phases. Usually, a soft material is viewed as a system with low anisotropy, K, (which results in a small coercivity, $H_C$) with a large saturation magnetization, $M_S$ (see Fig. 1) (see Table 2 for a summary of $M_S$ for a few selected soft magnetic materials). On the other hand, a hard material is taken as a material with a large K and a moderate $M_S$ (see Fig. 1) (see Table 3 for a summary of the properties of a few selected hard magnetic materials). In exchange coupled hard-soft thin film systems the magnetization switching behavior and, thus, the hysteresis loops, has been demonstrated to depend strongly on the dimensions of the soft phase (in thin films, the thickness of the soft layer, $t_{soft}$) [41,218–220]. For a thin $t_{soft}$ there is a critical thickness below which the soft phase is rigidly coupled to the hard phase, and the two phases reverse at the same nucleation field, $H_N$, resulting in a rectangular hysteresis loop (see Fig. 1); in this case, the system is considered as completely exchange coupled. In contrast, for thicker soft layers, the soft phase nucleates the reversal at significantly lower fields and the switching is characterized by an inhomogeneous reversal, and the system is usually called exchange-spring magnet (see Fig. 1). Although the value of $H_N$ depends on the material parameters of both the hard and soft phases, the critical $t_{soft}$ is found to be roughly twice the width of a domain wall, $\delta_H$, in the hard phase: $\delta_H = \pi(A_{hard}/K_{hard})^{1/2}$ (where $A_{hard}$ and $K_{hard}$ are the exchange stiffness and anisotropy constants, respectively, of the hard phase – see Table 3 for a few examples of $\delta_H$). Thus, this length scale should determine many of the physical properties of these systems, independently of sample geometry [41,223,224].

**Table 2**
Saturation magnetization, $\mu_0 M_S$, of some selected soft magnetic materials at room temperature [221,222].

| Material | $\mu_0 M_S$ (T) / (emu/g) |
|---|---|
| Fe | 2.15 / 218 |
| Co | 1.81 / 161 |
| $Fe_{65}Co_{35}$ | 2.45 / 240 |
| $Fe_{20}Ni_{80}$ (Permalloy) | 1.00 / 93 |
| $Fe_3O_4$ | 0.63 / 92 |
| $MnFe_2O_4$ | 0.50 / 80 |

**Table 3**
Summary of various magnetic properties of some selected hard magnetic materials at room temperature [221,222].

| Material | $\mu_0 M_S$ (T) / (emu/g) | $K$ (MJ/m$^3$) / (Merg/cm$^3$) | $\delta_H$ (nm) |
|---|---|---|---|
| $Fe_{14}Nd_2B$ (FeNdB) | 1.61 / 171 | 4.9 / 49 | 3.9 |
| $SmCo_5$ (SmCo) | 1.07 / 110 | 17.2 / 172 | 3.6 |
| $Sm_2Co_{17}$ | 1.25 / 120 | 3.3 / 33 | 10 |
| $CoFe_2O_4$ (Co-ferrite) | 0.56 / 80 | 0.18 / 1.8 | 13 |
| $SrFe_{12}O_{19}$ (Sr-ferrite) | 0.48 / 71 | 0.35 / 3.5 | 14 |
| $BaFe_{12}O_{19}$ (Ba-ferrite) | 0.48 / 72 | 0.33 / 3.3 | 14 |
| FePt | 1.43 / 75 | 6.6 / 66 | 3.7 |
| CoPt | 1.00 / 50 | 4.9 / 49 | 5.0 |

In the case of thin $t_{soft}$, assuming small values for the anisotropy of the soft-phase, $K_{soft} \sim 0$, when $t_{soft} \leq 2\delta_H$, the two phases are rigidly coupled and the system is characterized by the *average* magnetic properties of the two layers. Then in the thin film case the expected value for the nucleation field, $H_N$ can be expressed as: $H_N = 2(t_{hard}K_{hard}+t_{soft}K_{soft})/(t_{hard}M_{hard}+t_{soft}M_{soft})$ (where K, M and t refer to the anisotropy, saturation magnetization and thickness of the soft and hard phases, respectively) [41,223,224]. For general geometries $H_N$ becomes, $H_N = 2(f_{hard}K_{hard}+f_{soft}K_{soft})/(f_{hard}M_{hard}+f_{soft}M_{soft})$, where f is the volume fraction of the soft and hard phases with $f_{hard} = 1 - f_{soft}$. Thus, this simple relation also reveals that depending on the relative volume fraction of the soft and hard counterparts $H_N$ will be dominated by either the soft or the hard properties. It should be taken into account that in the case of core/shell nanoparticles f is a non-linear function of the shell thickness, where if we have a core of radius R and a shell of thickness t, then the volume fractions become $f_{shell} = V_{shell}/V_{Tot} = 1- R^3/(R+t)^3$ and $f_{core} = V_{core}/V_{Tot} = R^3/(R+t)^3$. Consequently, small changes in the shell thickness may imply large changes in the volume fractions. For example, for a nanoparticle with a core of R = 7 nm, increasing the shell from t = 1 nm to 3 nm implies going from a core dominated volume fraction ($f_{shell} \sim 0.33$) to a shell controlled volume fraction ($f_{shell} \sim 0.63$). This will have a strong influence on the coercivity of the core/shell nanoparticles.

For example, as can be seen in Fig. 5, assuming a hard material with K = 5 MJ/m$^3$ and $\mu_0 M_S$ = 1 T and a soft material with K = 0.05 MJ/m$^3$ and $\mu_0 M_S$ = 2 T, $H_N$ changes drastically both in the conventional and inverse structures even for small changes in shell thickness. Moreover, in the strong exchange coupling case, if we assume perfect square loops the overall remanent magnetization, $M_R$, of the composite will be given by $M_R = f_{hard}M_{hard} + f_{soft}M_{soft}$. The figure of merit of a permanent magnet is given by the energy product, $(BH)_{max}$, i.e., maximum amount of magnetic energy stored in a magnet. This is an energy density and it is equivalent to the area of the largest rectangle that can be inscribed under second quadrant of the B-H (magnetic induction-magnetic field) hysteresis curve[225]. In the case of strongly exchange coupling this becomes $(BH)_{max} = \mu_o H_N M_R/2$ (for $H_N > M_R/2$), where $\mu_o$ is the permeability of vacuum [41,223,224]. Note that although $\delta_H$ of most typical hard magnets is relatively small (e.g., ~4-5 nm for $SmCo_5$, FePt or FeNdB or ~14 nm for Ba-ferrite see Table 3), $2\delta_H > 8-10$ nm is often larger than the typical sizes involved in core/shell nanoparticles. Consequently, it should be expected that many systems fall in the strongly exchange coupled category, with either the shell thickness or core diameter of the soft phase smaller than $2\delta_H$.

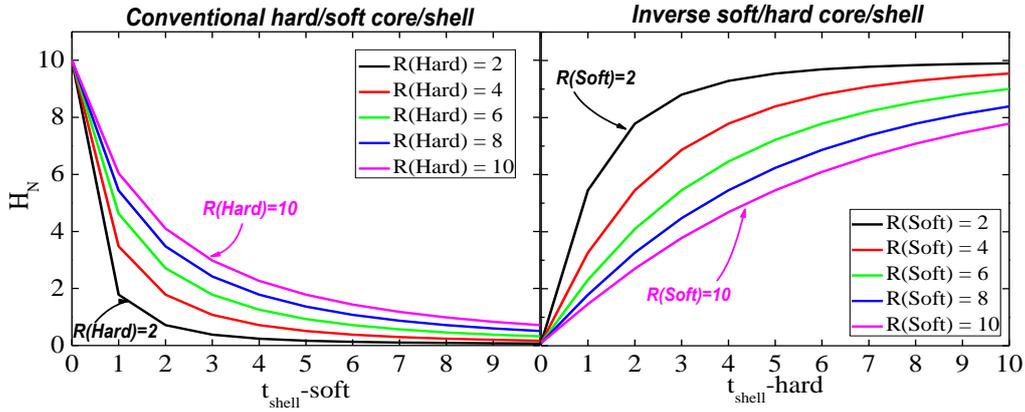

**Fig. 5.** Dependence of $H_N$ on the shell thickness for different core radii for conventional and inverse core/shell systems.

When $t_{soft} > 2\delta_H$, the soft layer nucleates the reversal at fields well below those of the hard layer. Under the assumption that the hard layer is perfectly rigid and $K_{soft} = 0$, solving for the magnetization of the soft layer with an applied field opposed to the hard layer it can be seen that the soft layer remains parallel to the hard layer for fields less than the nucleation field (or the exchange field of the soft phase, $H_{ex}$) given by $H_{ex} = \pi^2 A_{soft}/2M_{soft}(t_{soft})^2$. Once the magnetic field exceeds $H_{ex}$ magnetic reversal proceeds via a twist of the magnetization in the soft layer. This occurs because the soft layer is strongly pinned at the interface while away from the interface it is free to follow the external field. For $H > H_{ex}$, the spins in the soft layer exhibit continuous rotation, as in a magnetic domain wall, with the angle of rotation increasing with increasing distance from the hard layers. Such magnets exhibit reversible demagnetization curves since the soft layers rotate back into alignment with the hard phase if the reverse field is removed [41,223,224]. This reversal process is often referred to as an exchange-spring process by analogy with the elastic motion of a mechanical spring.

The above discussion neglects the effects of interparticle exchange and dipolar interactions. In particular, as discussed by Sebt et al., if core/shell nanoparticles are in physical contact with each other the exchange interactions could play an important role, where for example the coupling between two hard cores through their respective soft shells could lead to cooperative reversal and more complex effects such as random anisotropy effects [226]. Similarly, dipolar effects could also result in changes in the magnetic properties [226].

The main features described above, based on thin film bilayers or bulk nanocomposites, should be valid for core/shell nanoparticles. However, the particular morphology of the core/shell nanoparticles could lead to specific properties. In particular, contrary to thin film systems, the order of the layer in a conventional hard/soft or inverted soft/hard structure could play an important role in, for example, the reversal modes of the soft phase. For example in inverse soft/hard nanoparticles a new type of reversal mode (incoherent bulging) has been predicted [221]. However, in spite of the potential strong role of the shape, there are only few theoretical studies devoted to hard-soft core/shell nanoparticles [105,107,226–231]. Some simulations show that for hard/soft core/shell nanoparticles the evolution of $H_N$ on diverse parameters (e.g., shell thickness) follows quite closely the results expected from simple geometrical arguments (i.e., $f_{hard}/f_{soft}$) [231]. However, certain of these studies reveal, for example, large soft grains (in an inverse soft/hard structure) result in complex reversal or that the reduction of the interface coupling may enhance the spring-magnet type switching [228–230]. A recent systematic study of the effect of the hard/soft ratio for different nanoparticle sizes clearly evidences its important role on the magnetic properties (see Fig. 6a) [230]. In particular, it is shown that for each nanoparticle size the $(BH)_{max}$ is optimized for different hard/soft ratios (see Fig. 6b), where for certain nanoparticle sizes very large $(BH)_{max}$ could be obtained with relatively small hard phase counterparts [230]. Interestingly, in the case of the SmCo-Fe system, using periodic boundary conditions, the conventional SmCo/Fe core/shell nanoparticles seem to have improved properties with respect to the inverse Fe/SmCo ones [230].

Moreover, some studies on spherical soft magnetic inclusions embedded in a hard matrix, hard/soft nanoparticles embedded in a hard matrix, soft/hard nanoparticles embedded in a soft matrix, or hard inclusions in a soft matrix can be found in the literature [224,232–240]. The results show that the magnetic properties are strongly influenced by the dimensions and shape of the inclusions [224,232–234,240]. Moreover, due to the difference in magnetization between the hard and soft phases, dipolar interactions may affect the magnetic properties of the composites [235,236]. Interestingly, studies comparing spherical and planar soft inclusions in a hard matrix have determined that spherical inclusions are more effective in enhancing the magnetic properties of the nanocomposites [232,240].

When analyzing the magnetic properties observed experimentally for diverse hard-soft core/shell nanoparticles a spread of different behaviors can be found [91–95,98–100,102–104,106,107,110,111,114,115,121–123,125–132,136,140,141, 143, 175,241–295]. First, it should be pointed out that given that we are dealing with nanoparticles the core and shell sizes are rather small (i.e., usually smaller than $2\delta_H$), thus, most of the systems exhibit smooth hysteresis loops typical of strongly exchange coupled hard-soft counterparts. For conventional systems, i.e., when starting from a hard core, most systems report enhanced overall $M_S$ and reduced $H_C$ as soft layers are grown,[93,103,110,111,114,125,127–130,132,136,140,190,242–246,248,263, 271,275,276,286,288,289] as expected from the dependence of $H_N$ on the amount of soft phase (see Fig. 7).

When starting form a soft core (i.e., inverse structures) the opposite effects are found. Namely, $M_S$ decreases and $H_C$ rises as the amount of hard phase is increased [106,107,249,260,293]. However, in certain cases, if the difference in $M_S$ between the counterparts is small (e.g., hard Co-ferrite and soft Mn-ferrite both with $M_S \sim 80$ Am$^2$/kg (80 emu/g) at room temperature), although $M_S$ remains rather constant with the relative amount of hard and soft phase, $H_C$ does evolve according to the hard/soft volume ratio [101,102,143,251–253,262,292]. There exist some systematic studies fixing the core size and methodically increasing the shell size both for conventional hard/soft and inverse soft/hard systems [93,99,101,102,114,125,127–129,132,190,243,248,249,253,263]. The evolution of $H_C$, decreasing roughly linearly as the volume fraction of soft counterpart is increased (see Fig. 8), is consistent with the above simple theoretical arguments.

In some cases, two-stage loops, typical of spring-magnets, have also been observed [93–95,98,114,115,131,190,242,245,261,263,264,266,268,269,271,278,279,282–285,289,294,295] (see Fig. 9). As the thickness of a soft shell (conventional systems) or the diameter of a soft core (inverse systems) becomes very large the loop is no longer smooth and a constriction at low fields [93,114,115,190,245,261] is observed implying the formation of a pseudo-domain wall in the soft counterpart. However, several systems exhibit constricted loops even for rather small soft volumes. Thus, the origin of these loops does not always appear to be linked to the size increase of the soft counterpart to be larger than $2\delta_H$. One possible origin of the effect is that the two phases are only weakly coupled (e.g., due to exceedingly disordered interface). This would imply that the overall magnetic behavior would be a simple superposition of the properties of each of the counterparts. Another tentative cause of the constricted loops can be found in the change in microstructure of the systems when processed at high temperatures [98,264,278]. Namely, the original core/shell nanoparticles may coalesce at high temperatures forming a structure of hard nanoparticles embedded in a soft matrix. In this case the designation of the "size" of the soft phase may be less well-defined, thus the system may behave as a spring-magnet rather than a strongly exchange coupled system. Nevertheless, other origins, such as broad particle size distributions, are also possible. Interestingly, in some cases when increasing the size of the soft phase beyond a certain thickness a change in evolution of the magnetic properties (e.g., $H_C$), without any apparent transition, rather than a transition associated to each of the phases (see Fig. 10) [102]. Moreover since $T_B$ is controlled by K and V, the ratio of hard/soft volumes determines $T_B$ (see Fig. 10). change of the loop shape, has been observed [127,128]. This probably indicates that the transition between the strong exchange coupled regime and the spring-magnet one is not a sharp one but rather occurs gradually.

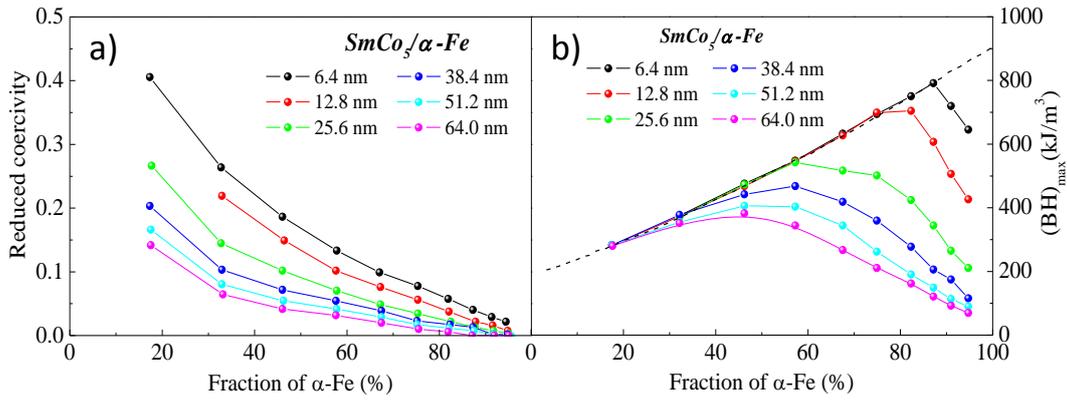

**Fig. 6.** Simulated dependence of the (a) coercivity and (b) energy product, $(BH)_{max}$ on the soft phase fraction for cubic SmCo/Fe core/shell nanoparticles overall lateral size L (see legend) [230].
*Source*: Reprinted/adapted figure with permission from Ref.[230] .
© 2013, by IEEE

Note that apart from the shell-shell contact (i.e., interparticle exchange interactions) dipolar interactions between particles can also play a dominant role in the magnetic properties [101]. In analogy to single phase nanoparticles [296–299] as they come closer to each other (i.e., stronger dipolar interactions) the magnetic properties and, in particular, the blocking temperature (see below) or the coercivity can be strongly affected.

Besides $M_S$ and $H_C$, the core/shell morphology also affects the blocking temperature, $T_B$. Due to their small volume magnetic nanoparticles are strongly affected by the thermal energy, where if the magnetic anisotropy energy, $KV$, where V is the volume of magnetic phase, is exceedingly small the thermal energy makes the spins fluctuate coherently giving rise to a state with zero net magnetization and vanishing $H_C$, i.e., superparamagnetism. If the temperature is lowered below a certain threshold the thermal energy is no longer sufficiently strong to make the spins fluctuate and the systems is said to be blocked, i.e., in a ferromagnetic-like state. The transition temperature between the blocked and superparamagnetic states is called the blocking temperature, $T_B$, and depends on the volume of the nanoparticle, its anisotropy and the characteristic measuring time of the technique, $\tau_m$ : $T_B = KV/(k_B \ln(\tau_m/\tau_o))$, where $k_B$ is the Boltzmann constant and $\tau_o$ is a time constant characteristic of the material (i.e.,

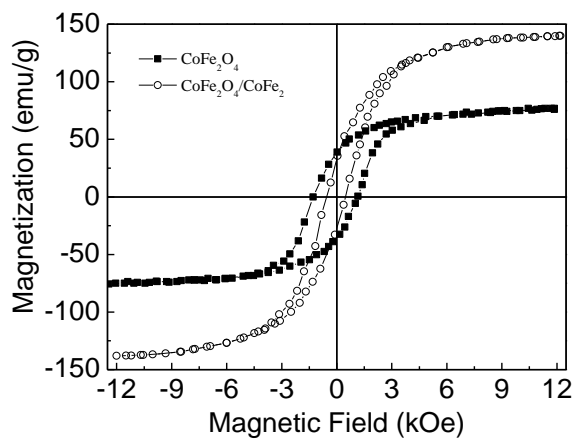

**Fig. 7.** Room temperature hysteresis loop for single phase $CoFe_2O_4$ nanoparticles and conventional hard/soft $CoFe_2O_4$/$CoFe_2$ core/shell nanoparticles [127].
*Source*: Reprinted/adapted figure with permission from Ref. [127].
© 2013, by Elsevier

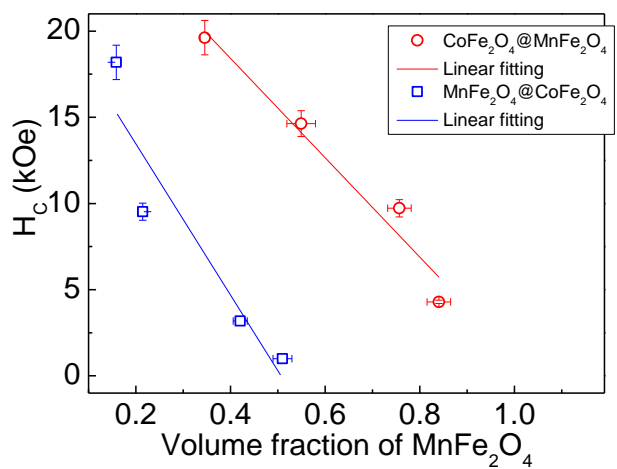

**Fig. 8.** Dependence of the coercivity, $H_C$, of bi-magnetic conventional $CoFe_2O_4$/$MnFe_2O_4$ and inverse $MnFe_2O_4$/$CoFe_2O_4$ core/shell nanoparticles on the volume fraction of the magnetically soft phase, $MnFe_2O_4$ [102].
*Source*: Reprinted/adapted figure with permission from Ref. [102].
© 2012, by the American Chemical Society.

the reversal attempt time, usually in the $10^{-12}$-$10^{-9}$ s range) [300]. For superconducting quantum interference device (SQUID) measurements, with $\tau_m \sim 100$ s, this becomes the well-known formula $T_B = KV/25k_B$. For typical temperature dependence of the magnetization measurements, M(T), the blocking temperature is usually taken as roughly the maximum of the zero field cooled magnetization. In the particular case of exchange coupled hard-soft core/shell nanoparticles the M(T) curve should exhibit a single

Concerning the role of the inverse vs. the conventional structure, although in the literature there are many examples of both arrangements, there are only a few studies which specifically compare both morphologies for a given set of hard and soft materials in the same conditions. Some examples are $CoFe_2O_4$-$ZnFe_2O_4$, $CoFe_2O_4$-$MnFe_2O_4$, $CoFe_2O_4$-$NiFe_2O_4$, $La_{2/3}Ca_{1/3}MnO_4$-$Sr_2FeMoO_6$ and $Fe_3O_4$-$Mn_3O_4$ [101,102,107,284,301]. At first glance, the results for the conventional and inverse morphologies are rather similar with no essential differences between both morphologies [101,102,107,284,301]. However, more accurate studies based on $CoFe_2O_4$-$MnFe_2O_4$ show that in conventional and inverse nanoparticles, with similar hard-soft volume ratio, $H_C$, is considerably larger in the conventional system than in the inverse one [102]. Unfortunately, since $M_S$ of both counterparts is very similar it is difficult to assess the role of the morphology in controlling the energy product. More systematic experimental and theoretical studies would be desirable to fully understand the differences between conventional and inverse systems.

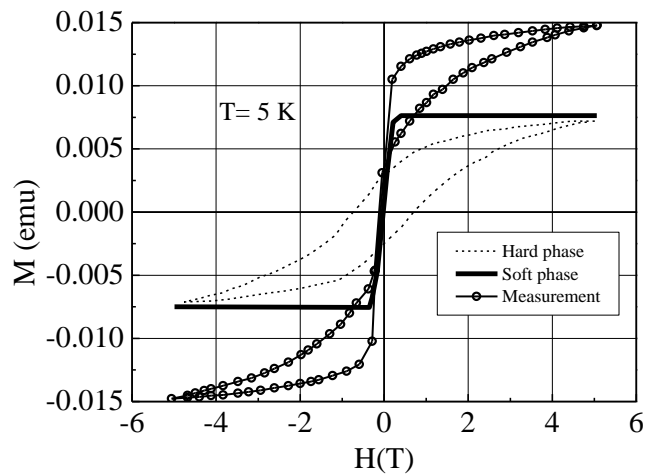

**Fig. 9.** Hysteresis loop (5 K) of FePt/iron oxide core/shell nanoparticles after annealing at 550 °C for 30 min [95].
*Source*: Reprinted/adapted figure with permission from Ref. [95].
© 2005, by the American Chemical Society.

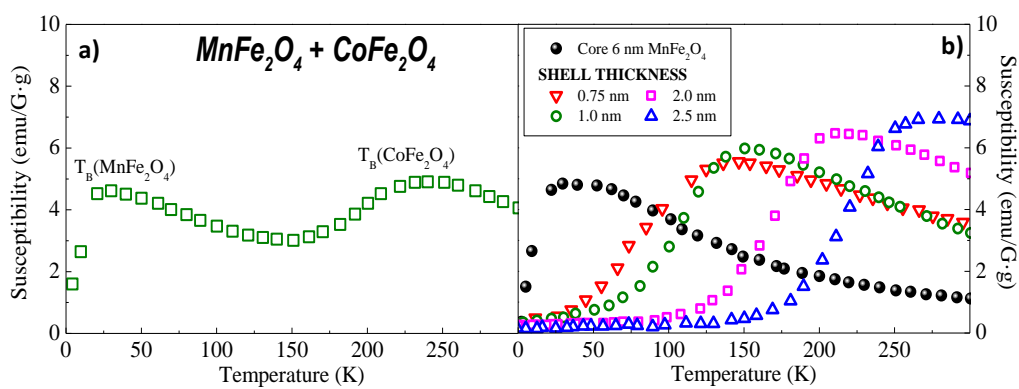

**Fig. 10.** Temperature dependence of susceptibility (H = 100 Oe) for (a) physically mixed $CoFe_2O_4$ and $MnFe_2O_4$ nanoparticles and (b) for $MnFe_2O_4/CoFe_2O_4$ core/shell nanoparticles for a fixed 6 nm $MnFe_2O_4$ core and different $CoFe_2O_4$ shells [102].
*Source*: Reprinted/adapted figure with permission from Ref. [102].
© 2012, by the American Chemical Society.

The hard-soft exchange coupling between core and shell has been addressed using more advanced characterization approaches such as ΔM-plots [302], first order reversal curves (FORC) [46,295,303], small angle neutron scattering (SANS) [199,304]or element-specific X-ray magnetic circular dichroism (XMCD) [305,306].

To investigate the nature of the interactions in magnetic systems, a procedure known as ΔM plot is often used [302]. ΔM is defined as $\Delta M = M_d(H)/M_r(\infty) - (1 - 2 M_r(H)/M_r(\infty))$; where $M_r(H)$ is the isothermal remanent magnetization, $M_r(\infty)$ is the remanent magnetization after fully saturating the sample and $M_d(H)$ is the demagnetization remanence. Usually, positive ΔM is interpreted as magnetizing exchange interactions, while negative ΔM corresponds to demagnetizing-like magnetostatic (dipolar) interactions [302]. In the case of hard-soft core/shell nanoparticles it has been observed that usually positive ΔM dominates the ΔM-plots, although the details of the shape of the ΔM-plots depend on the exact morphology of the samples [91,93,128,132,244,273]. Moreover, it has to be taken into account the interparticle dipolar and exchange (if they are in contact) may strongly influence the ΔM-plots. However as can be seen in Fig. 11a increasing the hard-soft contact area from a heterodimer morphology to a core/shell structure increases significantly the positive ΔM contribution as expected from the increase hard-soft exchange coupling [93]. However, it should be taken into account that multidomain particles, high order anisotropies or broad particle size distributions may complicate the interpretation of the ΔM-plots.

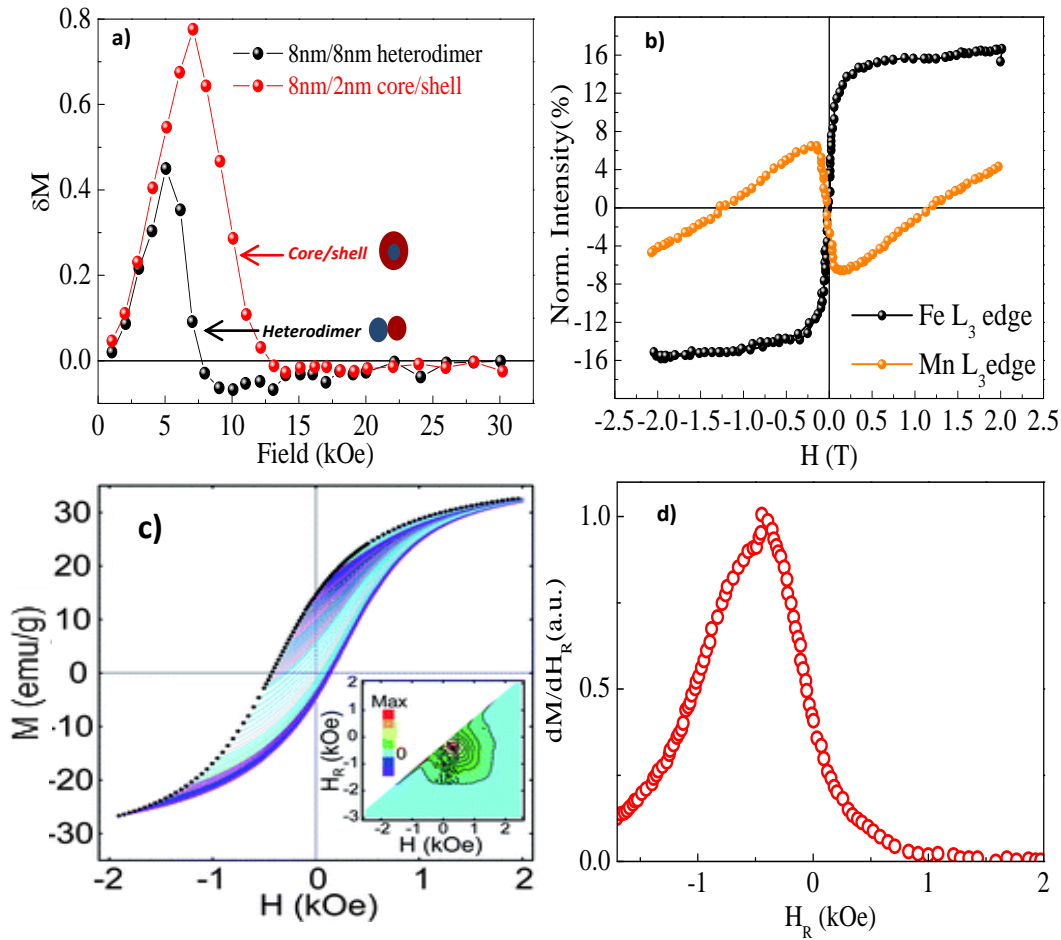

**Fig. 11.** (a) $\Delta M$ curves for a 8 nm/2 nm core/shell and 8 nm/8 nm heterodimer FePt/Fe$_3$O$_4$ nanoparticles after annealing at 650 °C for 1 h [93]; (b) Element-specific XMCD hysteresis loop at the Fe- and Mn-L-edges for Fe$_3$O$_4$/Mn$_3$O$_4$ core/shell nanoparticles [107]; (c) Families of FORCs for a MnFe$_2$O$_4$/FeMn$_2$O$_4$ core/shell nanoparticle, with the corresponding FORC distributions plotted in (H, H$_R$) coordinates shown as inset; (d) FORC switching field distribution of the FORC diagram shown in (c) [105].
*Source*: Panel (a) was reprinted with permission from Ref. [93].
© 2009, by American Institute of Physics.
Panel (b) was reprinted with permission from Ref. [107].
© 2013, by Nature.
Panels (c) and (d) were reprinted with permission from Ref. [105].
© 2012, by the Royal Chemical Society.

Another more advanced approach to evaluate the interactions is the FORC analysis. To obtain a FORC fingerprint, after positive saturation the applied field is reduced to a given reversal field, H$_R$. From this reversal field the magnetization is then measured back towards positive saturation, thereby tracing out a single FORC. This process is repeated for a series of decreasing reversal fields, thus filling the interior of the major hysteresis loop, which can be seen as the outer boundary of the family of FORCs (see Fig. 11c). The FORC distribution is then defined as a mixed second order derivative of the normalized magnetization:
$\rho(H, H_R) = -1/2 \, \partial^2 (M(H, H_R)/M_S)/\partial H \partial H_R$ which is then plotted against (H, H$_R$) coordinates on a contour map (see inset in Fig. 11c). For a given reversal field, H$_R$, the magnetization is measured for increasing applied fields, H, and therefore H ≥ $H_R$ by design. Following the measurement procedure the FORC distribution is read in a "top-down" fashion and from left to right for a particular reversal field. The FORC distribution provides a useful "fingerprint" of the reversal mechanism by mapping out, in (H, $H_R$) coordinates, only the irreversible switching processes. It is often useful to have a one-dimensional visualization of the irreversibility by projecting the FORC distribution onto the $H_R$-axis. This is equivalent to an integration over the applied field H:

$$\int [\partial^2 M(H,H_R)/\partial H \partial H_R] \, dH = \partial M(H_R)/\partial H_R$$

and is termed a FORC-switching field distribution (see Fig. 11d). Importantly, similar to ΔM-plots, FORC is also influenced by interparticle dipolar and exchange interactions. Studies of inverse soft/hard nanoparticles show that although FORC distributions do not show any well-defined characteristic features, the switching field distributions give some evidence of the coupling [105,106]. For example, as can be seen in Fig. 11d, the switching field distribution of $MnFe_2O_4/FeMn_2O_4$ nanoparticles is noticeably asymmetric, evidencing a tail towards high fields. This has been attributed to the presence of exchange coupled soft and hard phases in the nanoparticles [105].

In SANS the measured intensity, I, which is plotted as a function of scattering wavevector, Q, provides detailed information regarding the structural and magnetic morphologies in the micrometer to subnanometer length scale. Importantly, SANS probes magnetism, both in magnitude and spatial distribution as a function of applied magnetic field or temperature. Low temperature SANS measurements of soft/hard $Fe_3O_4/\gamma$-$Mn_2O_3$ nanoparticles show that the magnetization of the core and the shell switches coherently, as a single moment. This clearly evidences the strong exchange coupling between the two phases [199].

In XMCD the difference in X-ray absorption for left and right circularly polarized light is measured as a function of photon energy. Importantly, if the photon energy matches the absorption edge of a particular atom, information can be obtained on the magnetic properties of this specific atom, such as its spin and orbital magnetic moment. Thus, if the hard and soft materials have different types of atoms in their structure then by tuning the photon energy to the absorption edge of each specific atom information on each individual layer can be obtained. Particularly, XMCD spectra at different applied fields (or generally the dependence of the maximum XMCD intensity at a specific absorption edge on the applied field) can give information on, for example, the coercivity of each counterpart. XMCD element selective hysteresis loops carried out at the Mn and Fe $L_{2,3}$-edge of $MnFe_2O_4/FeMn_2O_4$ or $Fe_3O_4/Mn_3O_4$ nanoparticles also give information on the nature of the coupling. Given the phases involved in the nanoparticle the Mn-edge loop gives predominant information on the hard shell ($FeMn_2O_4$ or $Mn_3O_4$) whereas the Fe-edge loop gives information on the Fe-rich soft core ($MnFe_2O_4$ or $Fe_3O_4$). The element selective hysteresis loops show that the coercivities of the core and the shell are almost the same; however the approach to saturation of the hard shell is slower than the one of the soft core [105,107]. Interestingly, for $Fe_3O_4/Mn_3O_4$ nanoparticles it can be obviously observed that the XMCD signal for the Mn- and Fe-edge are inverted (see Fig. 11b). This clearly indicates an antiferromagnetic interface coupling between the core and the shell [107].

Another interesting basic property of hard-soft core/shell nanoparticles is that they have also been reported to exhibit exchange bias [94,105,107,122,247,259,263,279,307]. Namely, the hysteresis loops exhibit a loop shift in the field axis, $H_E$. Although exchange bias is a well-known effect of the exchange coupling between AFM and FM bilayers and core/shell nanoparticles [33–37], hard-soft bilayers (either in direct exchange contact or separated by non-magnetic layers) have also been reported to exhibit loop shifts [308–313]. In the case of hard-soft nanoparticles several systems have been reported to exhibit exchange bias. For example, there are some reports of exchange bias in the $CoFe_2O_4$-FeCo system, both in conventional and inverse structures [122,247,263]. Inverted $CoFe/CoFe_2O_4$ particles show rather small bias [122,247]. However, in the conventional $CoFe_2O_4$/FeCo case, a moderate bias has been reported. In this case $H_E$ depends inversely on the amount of soft phase [263], as expected from an interface effect [33]. Very large exchange bias has also been reported in $FePt/Fe_3O_4$ nanoparticles [94,283]. However, inverse $Fe/Fe_3O_4/FePt$ nanoparticles exhibit a small loop shift with a non-monotonic temperature dependence, which has been proposed to arise from the multishell character of the nanoparticles [279]. Finally core/shell nanoparticles based on Fe- and Mn-oxides have also been reported to exhibit loop shifts below the transition temperature of the Mn-oxide phases [105,107]. In the case of $MnFe_2O_4/FeMn_2O_4$ soft/hard nanoparticles, they have been shown to exhibit a small loop shift which also increases as the size of the soft core becomes smaller [105]. A related system exhibiting exchange bias is $Fe_3O_4$-$Mn_3O_4$, which shows loop shifts both in the conventional and the inverse cases [107]. Interestingly, for this system $H_E$ depends strongly on the cooling field, changing sign for sufficiently large cooling field. This has been interpreted as an indication of an antiferromagnetic coupling between the core and the shell [107].

Finally, note that core/shell nanoparticles with a soft or hard core and highly disordered or amorphous shells (which can have hard-like magnetic properties) can give rise to other more complex phenomena such as spin-glass like transitions, large irreversibility, large $H_E$, exchange bias training effects or strong dependence of $H_E$ on the cooling field [120,314]. Moreover, although bi-magnetic core/shell structures composed of soft/soft materials have been shown to also have appealing applications (e.g., in biosensors, hyperthermia, microwave absorption and magnetic resonance imaging) [89,315–317], they will not be discussed in this review. Similarly, nanoparticles with surface anisotropy effects [318–320] will not be considered in this review, even if they could be viewed as a purely "magnetic" (i.e., not structural) soft/hard core/shell configurations.

*2.3.2 Dynamic Magnetic Properties*

The magnetic dynamic properties describe basically the path and time that the magnetic moment will take to reach its final state after changing its direction or how the magnetization responds to an applied field which changes with time. When a magnetic material is subject to an ac-magnetic field, for high enough frequencies the magnetization cannot follow the applied field due to different kinds of magnetic losses. If we express the ac-field as $h_{ac}(t) = h_o e^{i\omega t}$, then the magnetic flux, B, is delayed by a phase angle $\delta$, thus $B = B_o e^{i(\omega t - \delta)}$ (where $h_0$ and $B_0$ are the intensity of the field and the flux, respectively, and $\omega$ is the angular frequency). Usually the losses are described by a complex permeability, $\mu = B/H$, $\mu = \mu' - i\mu''$, where $\mu'$ and $\mu''$ are related to the phase angle by $\tan \delta = \mu''/\mu'$. In bulk materials there are several different loss mechanisms, such as hysteresis loss, eddy currents, domain wall movement, and natural resonance [321]. Since usually at high frequencies the applied fields are rather small, the hysteresis loss is usually negligible. Interestingly, for nanostructured materials the eddy currents and domain wall movement mechanisms are not relevant either. Concerning eddy currents, these are induced currents in the material due to the ac-field which dissipate energy. However, when the skin depth (i.e., how deep the induced ac-currents penetrates in the material, given by $(\rho/\omega\mu)^{1/2}$, where $\rho$ is the resistivity [321]) is much larger than the size of the material the losses by eddy currents are negligible. Thus, since the skin depth of the field at high frequencies is rather large (usually in the range of µm), nanostructured samples are not affected by this loss mechanism. Moreover, below a certain size it is energetically not favorable to form domains in a magnetic material and the materials are in a single domain state [322]. Thus, the losses due to domain wall movement do not affect nanoparticles either. Consequently, the main loss mechanism in nanostructured materials is natural resonance. This resonance arises from the precessional motion of the magnetization due to the presence of a field. This field can be either external or intrinsic, for instance, arising from the magneto-crystal anisotropy. In simple terms, the magnetic field exerts a torque to the magnetization which due to the angular momentum causes the moments to precess. This process is well described by the Landau-Lifshitz-Gilbert equation
$dM/dt = -\gamma(M \times H) - \gamma\alpha/M_S[(M \times (M \times H))]$, where $\gamma$ is the gyromagnetic ratio and $\alpha$ is the damping parameter (assuming $\alpha \ll 1$) [323].

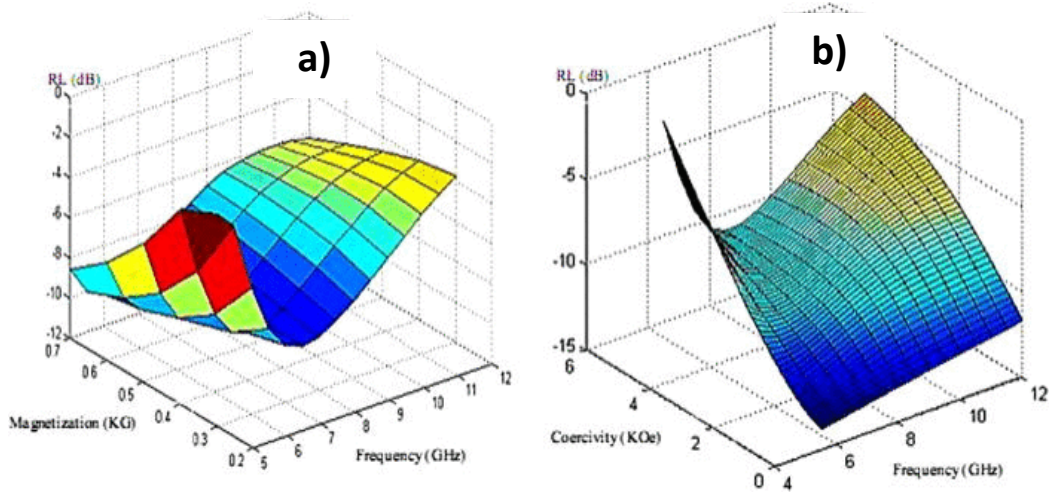

**Fig. 12.** Calculated frequency dependence of the reflection loss, $R_L$ on the (a) saturation magnetization and (b) the coercivity [339].
*Source*: Figure reprinted with permission from Ref. [339].
© 2013, by the American Institute of Physics.

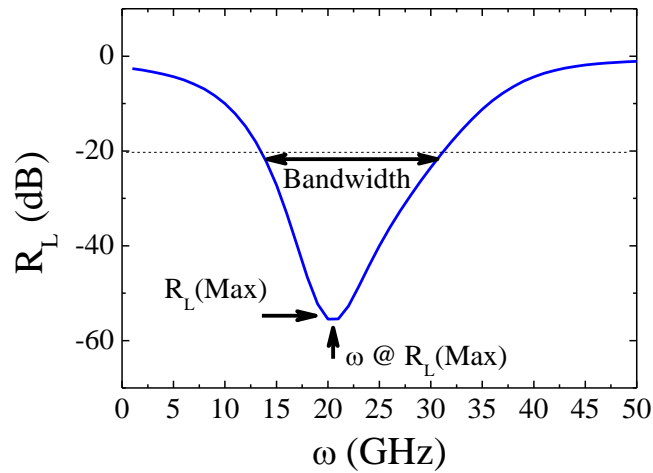

**Fig. 13**. Schematic representation of the frequency dependence of the reflection loss, $R_L$, where the main parameters are defined.

By solving the Landau-Lifshitz-Gilbert equation one obtains the resonance condition: $\omega = \gamma H_A$, where $H_A$ is the anisotropy field. In fact, $H_A$ should be substituted by an internal field $H_{int}$, which, apart from the anisotropy field, includes inter-particle dipolar interactions and the demagnetizing field. Importantly, if an oscillating magnetic field (for instance a microwave field) is applied at right angles to a static field at the same frequency as the precession (resonance condition) then the magnetization experiences an increase of the torque angle so that the energy from the microwave field is absorbed. This

A. López-Ortega, M. Estarder, G. Salazar-Alvarez, A.G. Roca, J. Nogués Physics Reports doi: 10.1016/j.physrep.2014.09.007.situation is commonly known as a ferromagnetic resonance (FMR) and can be obtained either by varying the frequency of the oscillating field or the strength of the static applied field. Nevertheless, this precession will not continue forever and in the absence of a driving field the magnetic moment will spiral until being parallel to the static field. This damping energy is dissipated to the lattice by means of the interaction of the spins with the lattice (creating in some cases spin waves) and is of great importance since it determines the time needed by the system to relax to the equilibrium state. In the absence of external field, the angular frequency will be governed by the magneto-crystal anisotropy, thus hard magnets will present higher resonance frequencies. For example while soft materials have their resonance frequencies in the low GHz range (e.g., $Ni_xZn_{1-x}$-ferrite - $\omega$ =1-10 GHz [324]), hard materials can have considerably larger resonance frequencies (e.g., Sr-ferrite - $\omega$ = 53 GHz [325]). However, the anisotropy field is not only governed by the anisotropy, K, but also by the saturation magnetization, i.e., $H_A = 2K/\mu_oM_S$. Thus, materials with moderate K but somewhat low $M_S$ can reach even larger resonance frequencies (e.g., $\varepsilon$-$Fe_2O_3$ - $\omega$ = 180 GHz [326]). Moreover, it has been shown that in certain materials by ion substitution or doping one can manipulate K and $M_S$ leading to the tunability of the resonance frequency. For example, in $\varepsilon$-$Fe_2O_3$ doped with other materials (e.g., $\varepsilon$-$Fe_{2-x}Z_xO_3$ with Z= Rh, Al or Ga) the resonance frequency has been shifted from 30 to 220 GHz [326–328].

However, despite their large resonance frequencies hard materials are not usually attractive for traditional microwave devices due to their low permeabilities [329], although novel high frequency applications using hard magnets are continuously emerging [330,331].

For a homogeneous magnetic material from the Landau-Lifshitz-Gilbert equation the complex permeability can be obtained:

$$\mu' = 1 + \{\gamma^2H_AM_S[\gamma^2H_A^2-\omega^2(1-\alpha^2)]\}/\{[\gamma^2H_A^2-\omega^2(1+\alpha^2)]^2+4\gamma^2H_A^2\omega^2\alpha^2\}$$
$$\mu'' = \{[\omega\alpha\gamma M_S[\gamma^2H_A^2+\omega^2(1+\alpha^2)]\}/\{[\gamma^2H_A^2-\omega^2(1+\alpha^2)]^2+4\gamma^2H_A^2\omega^2\alpha^2\}$$

[332]. Thus, although larger $H_A$ leads to high resonance frequencies, it also results in lower $\mu'$ and $\mu''$. Notably, for more complex materials such as core/shell nanoparticles embedded in a non-magnetic medium the analytical expressions of $\mu'$ and $\mu''$ can be rather complex [333–335]. Some of the limitations of hard ferrites stem from the so-called Snoek's limit, where the permeability and the resonance frequency are limited by the saturation magnetization, i.e., $\mu\omega_{FMR} = 2M_S\alpha/3$ [336]. However, hard-soft exchange coupling can in principle be used to tune, to a certain extent, the resonance frequency and $\mu'$ and $\mu''$ since it allows to control both $M_S$ and K of the composite material, where the effective K and $M_S$ are given by $K_{eff} = f_{soft}K_{soft} + (1-f_{soft})K_{hard}$ and $M_{S-eff} = f_{soft}M_{S-soft} + (1-f_{soft})M_{S-hard}$. Moreover, in certain cases, hard-soft composites may also act on the damping parameter, α, (since it can depend on surface and interaction effects [337,338]) giving an additional tuning parameter. This hard-soft approach is probably more versatile than the doping methods mentioned earlier since it allows a broader range of possible combinations and may, hence, overcome some of the limitations of microwave materials.

Another important aspect of the ferromagnetic resonance is that due to the magnetic losses the system absorbs part of the energy of the microwave field. Namely, if the incident microwave signal has a power $P_{in}$, when it interacts with a material part of incident power is absorbed by the material and part is reflected, i.e., $P_{in} = P_{Ref} + P_{Abs}$. Usually, to quantify the efficiency of the microwave absorption the results are given in terms of a reflexion loss $R_L = -10 \log (P_{Ref}/P_{in})$. Thus, large $R_L$ values imply stronger electromagnetic wave absorption. In the simplest case of normal wave incidence at the surface of a single layer material backed by a perfect conductor the reflection loss can be written as $R_L = 20 \log |Z_{in} - Z_0|/|Z_{in} + Z_0|$, where $Z_{in}$ and $Z_0$ are the input impedance and the impedance of free space, respectively [340].

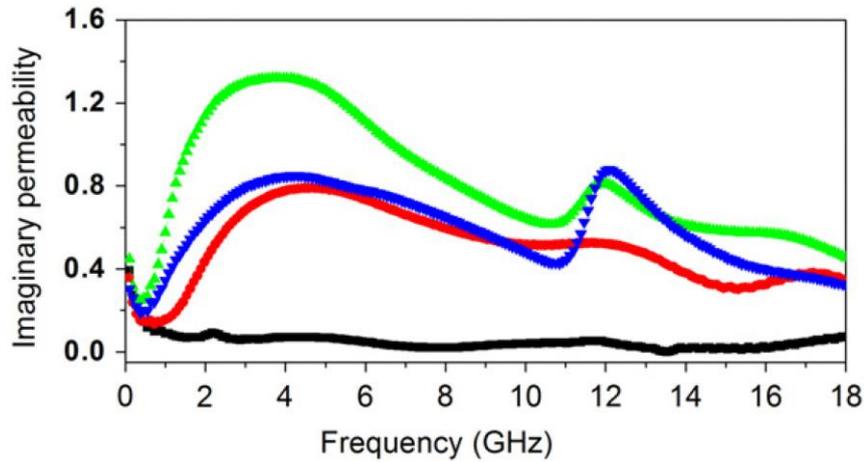

**Fig. 14.** Frequency dependence of the imaginary permeability, μ'', for $CoFe_2O_4/Co_3Fe_7$-Co core/shell nanoparticles with different hard/soft ratios [129].
*Source*: Figure reprinted with permission from Ref. [129].
© 2011, by the Institute of Physics.

$Z_0$ is a constant given by $(\mu_o/\varepsilon_o)^{1/2}$, where $\mu_o$ and $\varepsilon_o$ are permeability and permittivity of free space. On the other hand, $Z_{in} = (\mu_o\mu/\varepsilon_o\varepsilon)^{1/2} \tanh(j\omega d(\mu_o\mu\varepsilon_o\varepsilon)^{1/2})$, where $\varepsilon$ and $d$ are the permittivity and thickness of the absorbing material and $j$ the imaginary unit. Thus, taking into account that for many materials $\varepsilon$ is only weakly frequency dependent at high frequencies [340], the reflection loss depends mainly on the magnetic permeability in a complex way. Thus, since μ depends on $M_S$ and $K$, $R_L$ is also controlled by these parameters, as shown by theoretical calculations [339,341] (see Fig. 12).

Hence, in principle, $R_L$ should be moderately adjustable by hard-soft exchange coupling. Notably, due to the complex dependence of $R_L$ on the different parameters the microwave absorption efficiency and the relevant range of frequencies of a given material will depend on the combination of all parameters, i.e., the effective μ and ε of the overall system (i.e., magnetic material(s) and matrix) and its thickness. The dependence of $R_L$ on frequency usually exhibits a sharp minimum, where the key parameters are the frequency at which $R_L$ is maximum, the absorption intensity and the frequency range for which the absorption is larger than 20 dB, i.e., a 99% absorption (see Fig. 13).

Experimentally, a number of studies concerning the control of the microwave properties in different types of hard/soft systems (i.e., thin films, nanocomposites and core/shell nanoparticles) can be found in the literature [45,91,123,124,129,270,271,294,342–355]. Many studies show that μ' and μ'' depend on the ratio of hard/soft materials both for core/shell nanoparticles [129,342,345], nanocomposites [45,347–352] and bilayers [354].

However, as can be seen in Fig. 14 [129] for $CoFe_2O_4$/FeCo hard/soft core/shell nanoparticles, the dependence of μ'' on the hard/soft ratio can be non-monotonic. This is expected since μ' and μ'' are determined by both $M_S$ and $H_A$, which change simultaneously as the hard/soft ratio is modified. Concerning the resonance frequency, no systematic reports can be found for core/shell nanoparticles. However, in studies carried out in nanocomposites it has been observed that, indeed, the resonance conditions can be controlled by the hard-soft exchange coupling. For example, the hard/soft $Y_2Fe_{14}B$/FeB nanocomposite ribbons exhibit a fine control of the FMR frequency on the hard/soft coupling [45]. Consistent results are obtained from FMR measurements of hard/soft $Nd_2Fe_{14}B/\alpha$-Fe nanocomposite films at a fixed 135 GHz, which show that the field necessary to obtain the resonance conditions depends on the hard/soft ratio [353]. Similarly, both experiments and micromagnetic simulations of exchange coupled soft/hard thin film evidence that the exchange coupling is responsible for the changes observed in the FMR frequencies [354–356]. Interestingly, recent FMR results on $Fe_3O_4$(soft)-$Mn_3O_4$(hard) conventional and inverse core/shell nanoparticles with an antiferromagnetic interface coupling have shown that apart from the hard-soft exchange coupling, dipolar effects between the counterparts may also play an important role in the internal fields of the system, thus affecting the resonance conditions [107].

Due to increasing need for shielding of electromagnetic wave radiation at frequencies above 1 GHz, one of the most studied high frequency property of hard-soft systems is microwave absorption (i.e., reflection loss) – see Section 3.3. Studies on both hard-soft core/shell nanoparticles and nanocomposites have demonstrated that the main parameters of $R_L$ can be controlled by the hard/soft ratio [45,91,123,129,271,294,342–352].

## 3. Applications

*3.1 Permanent magnets*

Permanent magnets are used in widespread applications ranging from household items (such as cell phones), industrial applications (e.g., motors, alternators) to high-technology devices (like nano- and micro-electromechanical systems -NEMS-MEMS-, new types of recording media or magnetoresistive random access memories -MRAM-) [42,357–364]. Traditionally the search for new permanent magnets has been focused mainly on the search for materials with large anisotropies, mainly based on rare-earth elements, e.g., $SmCo_5$ or $Fe_{14}Nd_2B$ [42,357–365]. However, the ever increasing demand for permanent magnets has triggered a shortage of rare-earth raw materials resulting in a significant increase in price. This has generated renewed interest in alternative materials [364,366,367] such as hexagonal MnBi [368,369], $\varepsilon$-$Fe_2O_3$ [370,371], or $DO_{22}$ $Mn_{2-3}$Ga [372,373], $HfCo_7$ [374] or $L1_0$ MnAl [375,376] to mention a few. Moreover, in the 1980s–early 1990s a new type of permanent magnet material was proposed, namely, exchange-coupled hard-soft materials [223,377]. However, although it was not formally established yet, this concept was already used in the 1960s by Falk and Hooper to develop permanent magnets based on FeCo/Co-ferrite soft/hard, i.e., inverse, core/shell elongated particles [257,258]. This type of exchange-coupled hard-soft materials ideally combines the desirable properties of the hard (high $H_C$) and soft (large $M_S$) counterparts allowing, in principle, overcoming the limitations of conventional permanent magnets. However, conventional synthesis approaches for hard magnets (such as ball milling or sintering) offer poor control over the microstructure in these two-phase materials. Since the properties of the exchange-coupled magnets depend critically on the sizes of the hard and soft phases most of the basic research has been, in fact, carried out using thin films. Yet, thin films are not suitable for mass production. Thus, given the exquisite control wet chemistry offers over the dimensions of the different hard-soft core/shell nanoparticles, these are being pursued as a promising solution to create new rare-earth-free permanent magnets. Interestingly, nanophase materials offer additional advantages since they allow an easy control of the magnetic properties, which can lead to optimized properties for spring magnets. For example, $H_C$ of nanostructured magnets increases with the reduction of particle size going through a maximum at the single domain size [1–12]. On the other hand shape anisotropy of magnetic nanostructures could be exploited to induce large anisotropies, although the role of demagnetizing fields could be a limiting factor in magnet design [378].

The first patent using the concept of soft/hard core/shell nanoparticles was granted as early as 1964 to R.B. Falk and it dealt with the controlled oxidation of elongated FeCo nanoparticles to form a hard Co-ferrite shell. Maximum energy products in the range of 32 $kJ/m^3$ (4 MGOe) were reported [379]. The recent patents filed in this topic can be divided in three different categories: (i) the ones based on rare-earth hard magnets (FeNdB or SmCo) coupled to high moment soft magnets such as Fe or FeCo [47–58] (ii) nanoparticles of hard magnets based on precious metals (e.g., FePt or FePd) coupled to Fe, $Fe_3Pt$, $Fe_3O_4$ or FeCo [54,55,59–63]and (iii) novel structures based on rare-earth and precious-metals free hard magnets ($\varepsilon$-$Fe_2O_3$, MnAl, MnBi, Ba-ferrite or $Co_3C$), coupled to soft materials (e.g., Fe, FeCo, FeNi or $Co_2C$) [53–55,65–71]. Although most of the proposed systems are based on conventional structures, i.e., where the hard magnet is in the core, some notable exceptions are Fe/$\varepsilon$-$Fe_2O_3$ and Co/$Co_3C$ which, since the hard magnet is obtained by the surface treatment of the core, they have an inverse soft/hard structure [66,71]. The goal of type (i) systems is to reduce the amount of rare-earth necessary to fabricate magnets for diverse applications. Note that theoretically [224], energy products above those of pure rare-earth magnets could, in principle, be obtained in optimized exchange coupled hard-soft systems with less than 10% of hard magnet. Some patents in this group report experimental energy products in the range of 400-480 kJ/m$^3$ (50-60 MGOe) [47,48]. An alternative approach to reduce the amount of heavy rare earths (e.g., Tb or Dy) has been proposed by TDK Corporation in a series of patents based on bi-magnetic core/shell nanoparticles, where they propose a core with FeNdB type of material with no (or reduced amount) heavy rare earth ions covered by a shell also based on FeNdB material but containing heavy rare earth [376–379]. In this case both the core and the shell are hard materials; however the core tends to have slightly larger $M_S$ but lower $H_C$ than the shell. Given the high price of precious metals, the large scale industrial application of type (ii) composite magnets is less probable; however niche applications in small systems like MEMS could be envisioned. This type of systems has the advantage that compared to rare-earth magnets FePt-like nanoparticles are considerably easier to fabricate. Interestingly, energy products in the range of 720 kJ/m$^3$ (90 MGOe) have been reported for FePt/FeCo [61]. The goal of type (iii) particles is to develop novel rare-earth free permanent magnets. Most of the existing rare-earth free permanent magnets cannot really compete in energy product with NdFeB or SmCo, mainly because of their somewhat low $M_S$ (like $\varepsilon$-$Fe_2O_3$, MnAl or MnBi). However, theoretical calculations show that competitive energy product values above $(BH)_{max} \sim 400$ kJ/m$^3$ (50 MGOe) should be achievable in hard/soft core/shell nanoparticles using FeCo as the soft counterpart [68]. Finally note that most of the proposed materials are based on strongly exchange coupled core/shell nanoparticles forming a dense bulk-like material.

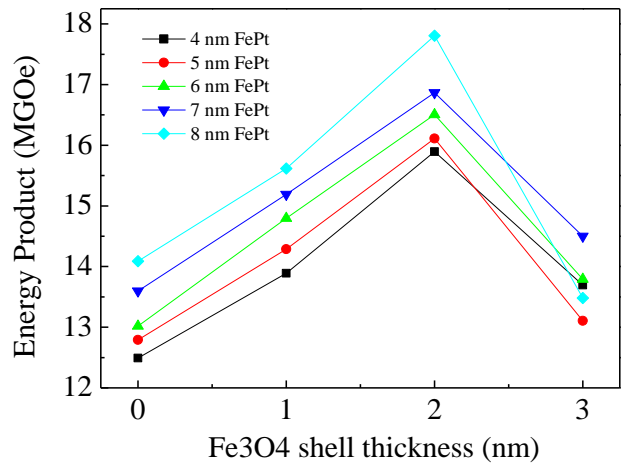

**Fig. 15.** Dependence of the energy product on the $Fe_3O_4$ shell thickness for different FePt core diameters [93].
*Source*: Reprinted/adapted figure with permission from Ref. [93].
© 2009, by American Institute of Physics.

From research articles the trend in the type of hard materials used is similar to the patented studies, i.e., rare-earth based materials (e.g., FeNdB, SmCo), FePt alloys (or equivalents)

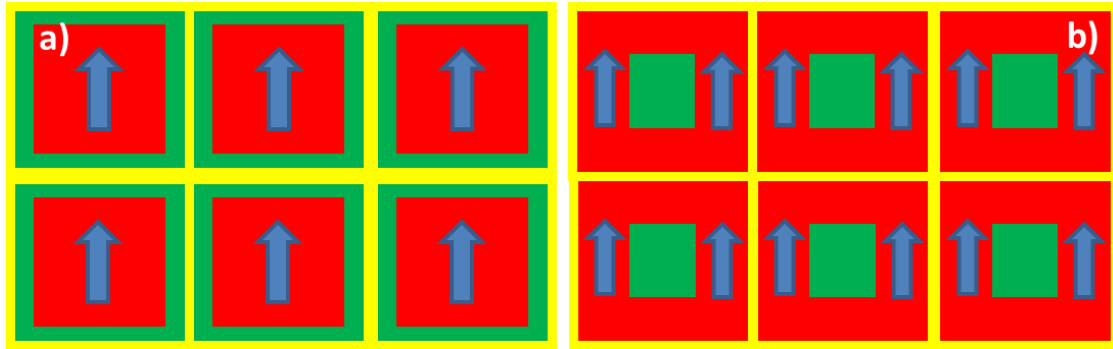

**Fig. 16.** Schematic representation of possible optimized structures for (a) hard/soft and (b) soft/hard core/shell nanoparticles. Shown in green, red with blue arrows and yellow are the soft, hard and non-magnetic counterparts, respectively [363]. (For interpretation of the references to color in this figure legend, the reader is referred to the web version of this article.)
*Source*: Reprinted/adapted figure with permission from Ref. [363]
© 2012, by Elsevier.

[93,104,110,111,114,115,136,190,241–244,281,282]. Interestingly, the main type of material studied for possible application as permanent magnets that is rare-earth and precious metal free, is $CoFe_2O_4$ [98,125–128,130,132,248,257,258]. In contrast to patented materials, virtually no reports of core/shell structures based on hexagonal ferrites (e.g., Sr- or Ba-ferrites) [273,289] or novel hard magnetic materials (like ε-$Fe_2O_3$, MnAl, MnBi, $Co_3C$, $Mn_3Ga$) can be found in the hard magnet academic literature. Most of the systematic studies concentrate mainly on the evolution of $M_S$ and $H_C$ with the core diameter and shell thickness [92,93,114,190], although usually no estimation of the energy product, $(BH)_{max}$, is given. However, values of 360-440 $kJ/m^3$ (45-55 MGOe) have been reported for Fe and FeCo coated NdFeB particles [241], 5-10 $kJ/m^3$ (0.5-1.2 MGOe) for $Fe_2Co$ shells on $CoFe_2O_4$ cores [125,130], 10 $kJ/m^3$ (1.2 MGOe) for Sr-ferrite/$Fe_3O_4$ [289] or 140 $kJ/m^3$ (18 MGOe) for FePt/$Fe_3Pt$ nanoparticles [93]. For this latter example it has been shown that $(BH)_{max}$ is optimized by relatively thin soft shells (2 nm thick) independent of the core diameter (in the range of 4-8 nm) (see Fig. 15) [93]. However, so far the values reported for $(BH)_{max}$ are on randomly oriented particles and little progress towards the optimization of $(BH)_{max}$ have been reported. Nevertheless some of these values are already approaching the results reported for bulk exchange coupled materials [380].

Despite the great potential of core/shell nanoparticles to develop permanent magnets with reduced (or zero) rare-earth materials, several challenges still remain [363]. First, to maximize $(BH)_{max}$ the remanence magnetization, $M_R$, should be high (close to $M_S$). It is well known that a random distribution of non-interacting single domain particles has $M_R/M_S = 0.5$. To increase $M_R/M_S$ the alignment of the easy axes of the particles is necessary. A perfect alignment of single domain particles would result in $M_R/M_S = 1$ [381]. Thus, processes rendering well aligned nanoparticles would be desirable. Second, as discussed earlier the size (i.e., thickness or diameter) of the soft phase should not be larger than $\delta_H$. However, care must be taken to avoid exceedingly large volume fractions of the soft phase. Namely, given the non-linear dependence of the volume fraction on the core diameter and the shell thickness even if the size of the soft phase is well below $\delta_H$, the volume fraction of the soft phase could be very large leading to a deterioration of the hard magnetic properties (see Fig. 5). Related to this, soft-soft shell-shell contact in hard/soft nanoparticles should probably be avoided, since the exchange coupling between soft areas may lead to the size of the soft counterpart to become larger than $\delta_H$ or might even promote reversal by nucleation and propagation of domain walls. Finally, nanoparticle compaction is also important to obtain strong permanent magnets. From these requirements perhaps favorable microstructures would be (i) hard cores with a thin soft shell protected by a thin non-magnetic shell (to avoid exchange coupling) with aligned magnetizations or (ii) small soft cores with a hard shell protected by a thin non-magnetic shell (see Fig. 16) [363].

*3.2 Recording media*

The race of magnetic recording towards areal bit densities beyond 1 Tbit/in$^2$ is being hampered by the so-called magnetic recording "trilemma" [382–385]. Namely, an increase in areal bit density implies a concomitant reduction of the bit size. To maintain both a sufficient signal-to-noise ratio (SNR) and high bit density, the volume of the individual grains that constitute a given bit must approach the superparamagnetic limit. While high uniaxial anisotropy ($K_u$) materials would dramatically improve thermal stability, the magnetic fields required for switching such a high $K_u$ bit would far exceed the capabilities of current write heads. Thus, to have viable recording media the thermal stability and SNR must be *simultaneously* balanced with the writtability of a given bit. To overcome this hurdle new types of recording media have been proposed such as exchange coupled composites [386], tilted anisotropy [387,388] and gradient anisotropy [389,390]. Nevertheless, these approaches also have their limits for sufficiently high densities. Thus, in the long term other kinds of media need to be designed. A promising candidate is the so-called 'patterned media', where rather than recording each magnetic bit over several hundred grains, each bit is recorded in individual magnetic dots [391–394]. A particular type of patterned media is based on the self-assembly of magnetic nanoparticles [395–398]. Yet, this approach has some inherent dilemma. Specifically, for the nanoparticles to be magnetically stable at room temperature they need to have a high $K_u$, this implies that the field available for conventional write-heads may not be sufficient to write the information. One possible way to circumvent this problem may be to use materials with lower $K_u$ (i.e., soft) and tune their effective anisotropy by coupling them to high $K_u$ (i.e., hard) [143] or by using antiferromagnets [399]. An alternative approach would be to reduce the effective anisotropy of high $K_u$ (i.e., hard) nanoparticles by coupling them with low $K_u$ (i.e., soft) materials, for example in a core/shell structure, similarly to what has been proposed for continuous thin film media [386,389,390]. In these cases, the soft counterpart acts to reduce the switching field (allowing the write-head to write the information) while the hard part preserves thermal stability. In fact, recent micromagnetic simulations on patterned media have demonstrated that hard nanostructures embedded in soft shell hold very appealing properties in terms of reduction of the switching field without severe loss of the thermal stability (see Fig. 17) [400,401].

Although there has been some activity in the field of hard-soft core/shell nanoparticles for their use as recording media [64,72–79], the patents in this field discuss mainly potential materials rather than fully operational recording media. In fact the concept of using core/shell nanoparticles for recording is not new. Already in the 1960s soft $\gamma$-$Fe_2O_3$ nanoparticles surface doped with Co, leading to a hard $CoFe_2O_4$ surface layer (i.e., $\gamma$-$Fe_2O_3$/$CoFe_2O_4$ inverse soft/hard structure) were studied as potential materials for conventional recording media [402–404]. This type of structure allowed increasing $H_C$ of the nanoparticles while maintaining a uniaxial anisotropy (in contrast to the cubic anisotropy of pure $CoFe_2O_4$ particles). In turn, the increased $H_C$ allowed reducing the size of the recorded bits [403,404]. Patents on this type of structures can be found as late as the mid-1990s [76–78]. Interestingly, recent simulations on the use of core/shell nanoparticles for conventional recording have shown that this structure allows particle size reduction (which leads to improved SNR) without compromising thermal stability[231]. However, the current trend in hard-soft core/shell nanoparticles for recording is for single particle patterned media recording [64,72–77,79]. Similar to the case of permanent magnets the different materials proposed for recording are based on either rare-earth alloys (FeNdB or SmCo) [72,79], FePt [64,72–74] or hexagonal ferrites (Ba-ferrite) [75,79] as the hard counterpart. All the patented materials have a conventional hard/soft core/shell structure, where the soft materials are based on either metallic (Fe or Co) or oxide ($Fe_3O_4$ or $\gamma$-$Fe_2O_3$) materials. A notable exception is the material based on a hard FePt core and a FeRh soft shell [74]. In this case the FeRh shell exhibits a transition from AFM to soft-FM above room temperature. Thus, the proposed recording mechanism is based on the so-called heat assisted magnetic recording (HAMR [405]), where as the temperature is increased the AFM turns into soft-FM, which by exchange coupling reduces the coercivity of the hard FePt core allowing the read head to write the magnetic bit. Upon decreasing the temperature to room temperature the shell becomes again AFM and the core recovers its high $H_C$ and the strong thermal stability.

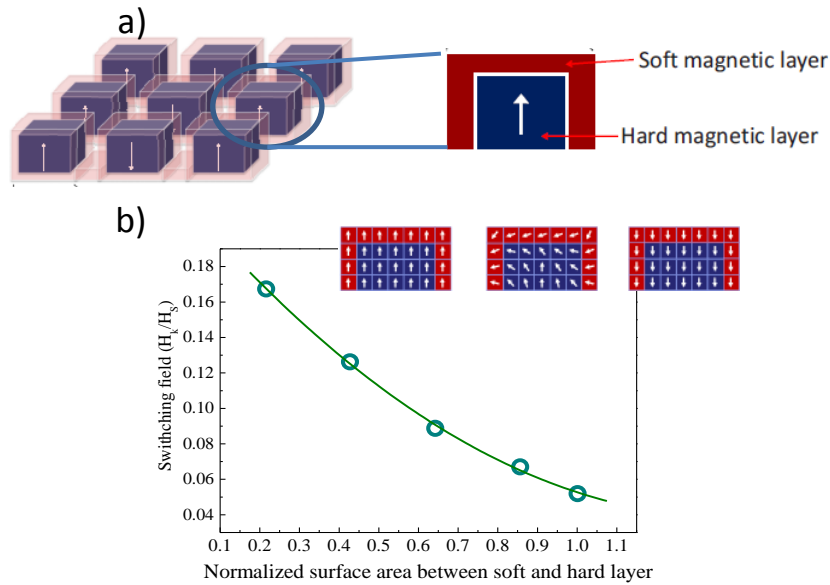

**Fig. 17.** (a) Nanostructured core/shell patterned media; (b) Micromagnetic simulation of the dependence of the switching field on the normalized surface area between the soft and hard layer. Shown in the inset is the reversal process of the structures [400].
*Source*: Reprinted/adapted figure with permission from Ref. [400]
© 2009, by the American Institute of Physics.

Interestingly, some of the patents demonstrate some degree of self-assembly for moderately large areas [64,72–74], which is a requirement for patterned media.
Although, there are no scholarly publications directly related to the use of hard-soft core/shell nanoparticles for magnetic recording media, in some studies the possible application of this morphology in future recording media applications is discussed [92,95,141,143,246]. Actually, some of the basic properties which could be appealing for the use of core/shell nanoparticles as recording media have already been experimentally demonstrated. For example, (i) it has been demonstrated that $T_B$ can be easily increased by adding a surface layer of a hard material on a soft nanoparticle (soft/hard approach) [102,143,250], (ii) it has been shown that the coercivity (i.e., switching field) of hard nanoparticles is reduced when coupled to soft shells (hard/soft approach) [102,190], (iii) the possibility to form self-assembled arrays of core/shell nanoparticles [92–94,99,102,103,122,190,263,406] or (iv) possible graded anisotropy in core/shell nanoparticles [105]. Nevertheless, other critical features like large, defect free, self-assemblies of hard-soft core/shell nanoparticles (i.e., in the same range as the ones shown for single phase magnetic nanoparticles [407]) have not yet been demonstrated. In fact, although some types of inorganic core/shell nanoparticles have been shown to form well-ordered arrays [21], since as the shell is grown on the core the overall morphology can slightly deteriorate, consequently the self-assembly becomes increasingly difficult, as demonstrated in Fig. 18 for FePt/$Fe_3O_4$ nanoparticles [93]. Note the deterioration of the self-assembly symmetry as the $Fe_3O_4$ shell is grown on the FePt nanoparticles. Thus, not only the quality of the seeds is important, but the morphology of the shell is also a critical parameter to obtain large defect-free self-assembles. Nevertheless, guided self-assembly, i.e., in confined areas or pre-patterned areas [408–410], could lead to less strict conditions for the self-assembly of the core/shell nanoparticles.

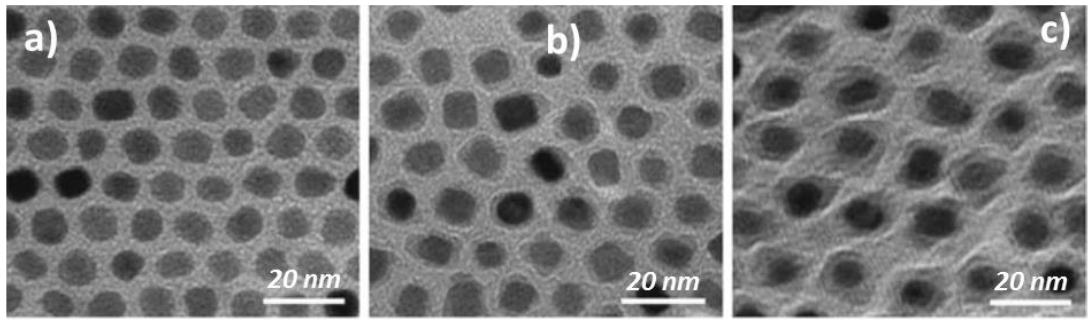

**Fig. 18.** TEM images of the as-synthesized (**a**) 7 nm FePt nanoparticles; (**b**) 7 nm FePt nanoparticles coated with 1 nm $Fe_3O_4$ shell; and (**c**) 7 nm FePt nanoparticles coated with 3 nm $Fe_3O_4$ shell [93].
*Source*: Reprinted/adapted figure with permission from Ref. [93].
© 2009, by American Institute of Physics.

In spite of the promise of core/shell structures for high density recording several challenges still remain. In fact, the conditions to use core/shell nanoparticles in recording applications are even more stringent than for permanent magnets. For example, if patterned media recording using core/shell nanoparticles should follow the principles of current magnetic recording then (i) all the nanoparticles should be single domain and exhibit out-of-plane magnetization, (ii) they should have a very narrow switching field distribution (linked, among other things, to the particle size distribution); (iii) the particles should not be in contact or should have a non-ferromagnetic protecting layer to avoid interparticle exchange coupling; (iv) dipolar fields should be small enough to avoid accidental recording of adjacent particles; (v) the coercivity needs to be smaller than the available writing field; (vi) the particles must be thermally stable and preferably with a large saturation magnetization; (vii) the 2D ordering should be defect free and preferably in square arrays. Remarkably, as discussed earlier, the latter condition remains one of the main hurdles for the development of recording media. Notably, another critical issue for self-assembly is that most of the large 2D ordered self-assemblies have been demonstrated with nanoparticles which are superparamagnetic at room temperature. Strongly magnetic nanoparticles tend to agglomerate or form non-ordered 2D arrays [411,412]. Other conditions [e.g., (v) and (vi)] may be easier to fulfill in hard-soft core/shell nanoparticles than for single phase nanoparticles.

Interestingly, progress in multilayered, onion type, nanoparticles [193,262,293,413,414] and AFM coupled core/shell nanoparticles [107] open new perspectives in the field of magnetic recording. In principle, the magnetic structure of this type of nanoparticles would allow for higher density recording where each nanoparticle could record more than two states, i.e., multilevel recording, similar to what has been proposed for thin film and lithographed structures [415–417]. However, there exist no detailed studies on this topic.

*3.3 Microwave absorption*

Microwave devices have been traditionally based on bulk and thin film ferrites [330,418]. However, in recent years it has been shown that nanocomposite materials, based on magnetic nanoparticles, can have certain advantages over conventional materials in specific applications [333,341,419–421]. Another driving force for the development of nanocomposite microwave materials is the increasing number of devices and communications standards (e.g., Wireless Local Area Network, WLAN) which are pushing the frequency to the Super High Frequency range (3-30 GHz) and millimeter wave range (30-300 GHz) which are above the resonance frequency of conventional soft ferrites [325,328,422]. Interestingly, with the boost of devices working at GHz frequencies (e.g., mobile phones, microwave communications or different types of radars) electromagnetic "pollution" can become a serious problem. Electromagnetic interference can be detrimental to many devices leading to, for example, poor communications. Thus, there is an increasing need for microwave absorbing materials which can attenuate unwanted electromagnetic signals. Most microwave absorbing materials are composed of magnetic loss powders such as ferrites or Co and dielectric loss materials such as carbon materials, metal oxides or polymers [22,423–428]. In simple terms, the perfect absorber should have a strong absorption in a wide frequency bandwidth (i.e., a large working frequency range), it should work at zero applied magnetic field and it should be thin and light weight [341].Thus, in recent years there has been an increase in the search for novel types of microwave absorbers and, in particular, in systems involving bi-magnetic hard-soft core/shell nanoparticles [80–82].

The patents discussing the use of bi-magnetic core/shell nanoparticles for microwave absorption show that having a core and a shell with dissimilar magnetic properties can enhance their performance [80–82]. Thus, although this field has not been as investigated as other aspects of hard-soft core/shell nanoparticles, the available results envisage a great potential for this application.

The first example is a material composed of a soft core (based on FeAlSi alloys) and a magnetically harder shell based on electroplated (FeCo or CoNi alloys), i.e. inverse soft/hard structure. The static magnetic properties show that the $H_C$ of the composite system is increased to $H_C > 150$ Oe. The high frequency results evidence that the presence of the shell can indeed improve the reflection loss properties of the material in the GHz range. Nevertheless, the absorbing power remains moderate [80]. More recently, Henning et al. have proposed, among different types of nanoparticles, the use of inverse soft/hard systems based on Co/CoFe$_2$O$_4$ as microwave absorbers. Unfortunately, they do not present any experimental data on this system [81]. Finally, in a recent patent, Imaoka et al. have proposed to use RE-Fe-N (e.g., RE = Sm or Nd) based materials as the hard counterpart in hard/soft core/shell microwave absorbers. They demonstrate that by using shells of different types of soft materials (e.g., Fe$_3$O$_4$, Ni-ferrite or Zn-ferrite) both the resonance frequency and the permeability can be tuned [82].

From the research articles on this topic it can be seen that hard-soft nanoparticles offer improved reflection loss characteristics with respect to single phase hard or soft nanoparticles. Diverse types of systems have been studied, based on different hard magnetic materials, e.g. CoFe$_2$O$_4$ [123,124,129], hexagonal ferrites (e.g., Sr-ferrite or Ba-ferrite) or more novel materials such as Fe$_{16}$N$_2$ [91,270,271,342–346]. Moreover, the patents on core/shell nanoparticles and the studies on nanocomposite materials (i.e., not core/shell) have also shown that other hard materials may be potentially interesting as reflection loss materials, e.g. FeNdN [82] and FeNdB [351]. The different core/shell systems have shown advances in diverse aspects of reflection loss: increase in $R_L$ [91,124,129,270,271,294,342,344–346] (see Fig. 19), higher working frequencies [124,271,294] or improved bandwidths [91,344,345].

Unfortunately, the prediction of the absorption properties is somewhat complex since it has to combine the magnetic and dielectric properties of not only both the core and the shell but also the matrix material. Moreover, the magnetic properties will strongly depend on the magnetic coupling and morphology, i.e., core diameter, shell thickness and the shape of the particles [419]. However, some simple predictions can be made. For example, although the relation between μ and $R_L$ is not linear, as a rule of thumb, the higher the resonance frequency of the material the larger the $R_L$ working frequency [429,430]. Similarly, one of the main goals when designing of microwave absorbing materials is reducing their thickness. From the definition of $R_L$ (section 2.3.2, Dynamic properties) it can be seen that there is an optimum thickness of the material where the $R_L$ is maximum (i.e., matching thickness). Under certain conditions, the matching thickness, d, of electromagnetic absorbing materials is closely associated with their imaginary permeability $\mu''$, $d \propto 1/\omega \mu''$, i.e., d is roughly inversely proportional to $\mu''$ [431]. Consequently, in first approximation, to decrease d, one effective approach is to increase its $\mu''$ [419]. Moreover, the design of the material will depend on the needs of the application. If one needs a narrowband of absorption, i.e., very high absorption at a single frequency, it is best to work at the matching thickness. However, if a broadband absorption is needed, i.e., absorption in a range of frequencies (rather than a single one), then it may be better not to work at the matching thickness. As can be seen in Fig. 19, off the matching thickness there is a loss in absolute $R_L$, however often there is an increase in bandwidth. Also as a first approximation the absolute value of $R_L$ can be increased if the ratio $\mu/\varepsilon$ is increased. Consequently, a simple procedure to improve the absorption is to increase the permeability [419].

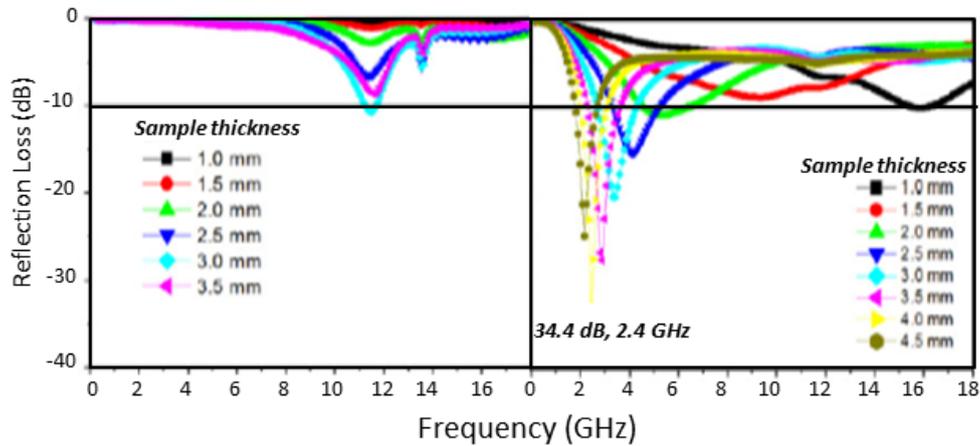

**Fig. 19.** Reflection loss spectra of (a) single phase CoFe$_2$O$_4$ nanoparticles and (b) CoFe$_2$O$_4$/FeCo hard/soft core/shell nanoparticles for different sample thicknesses [129].
*Source*: Reprinted/adapted figure with permission from Ref. [129].
© 2011, by the Institute of Physics.

Finally, it is worth emphasizing that since new high frequency, microwave and millimeter-wave, devices are continuously being developed [330,331], it should be expected that hard-soft core/shell nanoparticles may also play a role in these novel types of applications not only as microwave absorbing materials but as active components in, for example, circulators, isolators, phase shifters, filters or even metamaterials.

*3.4 Biomedical applications*

Biomedical applications are undeniably one of the most burgeoning fields in the use of magnetic nanoparticles, where many applications in very diverse aspects of biomedicine, not only for diagnosis but also for the treatment of diseases, are being developed [432,433]. In fact, the relevance of magnetic nanoparticles in biomedicine, e.g., in hyperthermia [434,435], drug delivery [436,437], biosensors [438,439], protein and cell manipulation (e.g., magnetic separation) [440,441], or magnetic resonance imaging (MRI) [442,443], among others, has already been widely demonstrated. In fact, the unique properties of nanoparticles make them excellent platforms to combine both therapeutic and diagnostic capabilities in a single entity, i.e., theranostics [444]. Similarly, magnetic nanoparticles are good building blocks for multifunctional core/shell nanoparticles for biomedical use, i.e., combining the magnetism (e.g., for hyperthermia) and another functionality (e.g., for optical imaging or to improve biocompatibility) [29,31]. However, from the magnetic perspective, many of the biomedical applications of magnetic nanoparticles rely solely on having a large $M_S$ and avoiding agglomeration to maximize the effects. Thus, most of these applications are based on the use of superparamagnetic nanoparticles [432,433]. In these cases there is no real need for hard-soft core/shell nanoparticles. However, in other applications the use of bi-magnetic core/shell nanoparticles can be beneficial. Actually, the use of bi-magnetic hard-soft nanoparticles for biomedical applications is still at its infancy.

Given the rather recent interest in bi-magnetic hard-soft core/shell nanoparticles for biomedical applications, only a few patents related to this topic have been filed so far [84–88]. Interestingly, the topics of the patents comprise different fields: hyperthermia [84,85], MRI imaging [86,87] and cell manipulation [88]. The first patent deals with conventional and inverse core/shell nanoparticles based on $CoFe_2O_4$ (hard ferrite) and diverse soft ferrites (e.g., $Fe_3O_4$ or $MnFe_2O_4$). It is found that the specific loss power of core/shell nanoparticles is better than both conventional materials and single phase nanoparticles of similar sizes. In particular, inverse $MnFe_2O_4/CoFe_2O_4$ nanoparticles are shown to have outstanding hyperthermia properties [84]. Concerning the nanoparticles for MRI imaging, they are based on hard $CoFe_2O_4$ cores capped with a soft ferrite (e.g., $ZnFe_2O_4$ or $MnFe_2O_4$). The nanoparticles are claimed to be an excellent $T_2$ contrast agent [86]. The last case is based on functionalized soft/hard nanoparticles designed to bind to specific cells. In the as-obtained state the particles present zero net moment (and consequently no agglomeration) thus they can easily attach to different cells. When a field is applied the particles become strongly magnetic and can be easily manipulated [88].

The academic research on biomedical applications is mainly focused on hyperthermia. Magnetic hyperthermia is based on the fact that nanoparticles when subjected to an alternating magnetic field produce heat [434,435]. Hence, if nanoparticles are in contact with an organic tissue the temperature increase produced by the nanoparticles can induce localized death of the targeted cells. Thus, hyperthermia has been proposed as potential novel targeted cancer treatment with higher efficacy and reduced side effects [434,435]. Three main mechanisms are involved in hyperthermia, Brownian relaxation, Néel relaxation and hysteresis loss, although for magnetically blocked nanoparticles only hysteresis loss prevails [445]. Hence, the specific loss power (SLP, a figure of merit of hyperthermia) of nanoparticles depends on material, size, composition and the frequency and intensity of the applied magnetic field [446–448]. Since heat dissipation depends on intrinsic material parameters such as $K_u$ and $M_S$, the combination of two phases in a core/shell structure can be advantageous to enhance the response of the material to the ac-fields, as shown by theoretical calculations [449,450]. Experimentally, it has been demonstrated that the combination of different magnetic soft/hard or hard/soft ferrite core/shell particles (e.g., $MnFe_2O_4/CoFe_2O_4$ or $CoFe_2O_4/MnFe_2O_4$) leads enhanced SLP values compared to single phase ferrite particles (see Fig. 20) [103,255,256]. Remarkably, inverse core/shell systems exhibit larger SLP values than conventional hard/soft structures (e.g., $SLP_{CoFe_2O_4/MnFe_2O_4}$ ~ 2300 W/g, $SLP_{MnFe_2O_4/CoFe_2O_4}$ ~ 3000 W/g) [103]. Moreover, further enhancement of the SLP values has been obtained in cubic soft/hard $Zn_{0.4}Fe_{2.6}O_4/CoFe_2O_4$ core/shell nanoparticles, with SLP values in excess of 10000 W/g. The decrease on the surface anisotropy and the size of the particles (60 nm) appear to be key factors for achieving SLP values 14 times higher than the spherical single component ferrite particles and 3 times than that of exchange-coupled spherical nanoparticles [255]. Apart from $CoFe_2O_4$ based systems, hard Sr-ferrite coupled to soft $\gamma$-$Fe_2O_3$ has also been proposed as a possible candidate for hyperthermia applications [269,451]. Another appealing core/shell system for hyperthermia is the one composed of Fe(soft)/FeC(hard) nanoparticles, which have revealed tunable magnetism as a function of the FeC shell. The results show that core/shell nanoparticles (with high $M_S$ and moderate $K_u$) exhibit superior properties to single phase Fe or FeC particles, with values of SLP up to 415 W/g [406]. Moreover, based on their large imaginary susceptibility (indicating large magnetic losses), $CoFe_2O_4$-$NiFe_2O_4$ nanoparticles have also been proposed as possible hyperthermia agents [284].

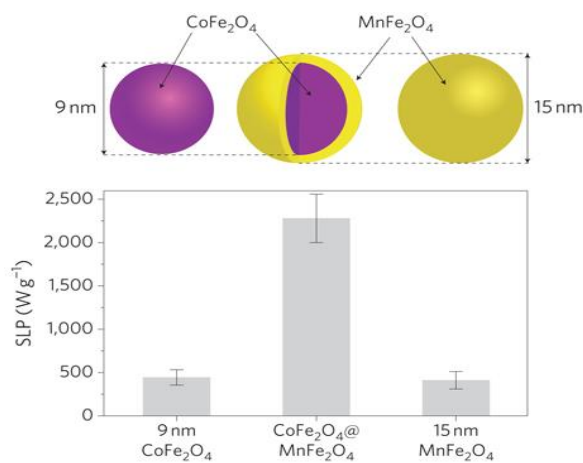

**Fig. 20**, Schematic representation of the $CoFe_2O_4$ seeds, the $CoFe_2O_4/MnFe_2O_4$ core/shell nanoparticles and a $MnFe_2O_4$ nanoparticle with the same total diameter (top) and their SLP values (bottom) [103].
*Source*: Figure reprinted with permission from Ref. [103].
© 2011, by Nature.

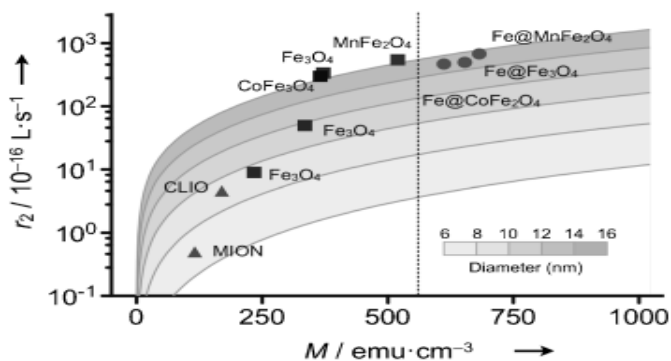

**Fig. 21.** $r_2$ relaxivity as a function of their M for different ferrofluids [100].
Source: Figure reprinted with permission from Ref. [100].
© 2011, by Wiley.

Another aspect of biomedical applications of core/shell nanoparticles studied experimentally is their use in MRI. Magnetic nanoparticles present several advantages compared to the conventional materials for certain MRI applications. For example, superparamagnetic iron oxide nanoparticles are known to decrease the transverse relaxation time, $T_2$, of the water protons when they are influenced by the nanoparticle dipole moment. This $T_2$ decrease leads to an increase of the negative contrast [442,443]. The relaxivity, i.e., the inverse of the relaxation time $r_2 = 1/T_2$ is proportional to the magnetic moment, thus, in a first approximation, increasing the magnetic moment of the magnetic nanoparticles is an efficient way to improve $T_2$- weighted imaging. However, as can be seen in Fig. 21 the correlation between $r_2$ and $M_S$ is not linear, thus other parameters (e.g., particle agglomeration) may influence the efficiency of magnetic nanoparticles for MRI. Consequently, there is an increasing interest in the study of different types of bi-magnetic core/shell nanoparticles for MRI [100,280,283,452], including soft/hard particles like $Fe/CoFe_2O_4$ which exhibit rather attractive $r_2$ values [100]. The improved MRI capabilities of core/shell nanoparticles make them also appealing for magnetic resonance based sensors such as diagnostic magnetic resonance [100].

Finally, contrary to the patents, there are nearly no systematic academic studies concerning ferrofluids for the manipulation of biochemical matter[292]. However, there is some work in developing diverse functionalization strategies of hard/soft particles like $CoFe_2O_4/Fe_3O_4$ or $FePt/Fe_3O_4$, which may be used in biomedical applications [90,94]

Although the use of hard-soft core/shell nanoparticles for hyperthermia is already rather well established, other biomedical uses of this type of systems have just started to emerge. However, the tunable magnetic properties of hard-soft core/shell nanoparticles will certainly lead to novel applications in the biomedical field, like for example in diverse types of biosensors. Nevertheless, one of the challenges for the *in-vivo* biomedical applications of magnetic nanoparticles is their possible cytotoxicity [453].

*3.5 Other Applications*

Apart from the conventional applications discussed above other more innovative uses of hard-soft core/shell nanoparticles can be envisioned.

For example, the miniaturization trend in many magnetic devices is driving the state of the art magnetic devices to the tens of nm dimensions [11,16,454–457]. Thus, to further push the limits of some magnetic devices core/shell nanoparticles could be an attractive alternative. In fact, the ability to measure magnetoresistance in single magnetic nanoparticles [458–460] is paving the way to future magnetotransport devices based on single core/shell nanoparticles. Actually, magnetoresistance measurements in agglomerates of core/shell nanoparticles have already demonstrated some potentially interesting effects. For example, $FePt/Fe_3O_4$ nanoparticles exhibit a tunneling magnetoresistance which changes sign with temperature [461]. Moreover, magnetoresistance measurements in the $La_{2/3}Ca_{1/3}MnO_4$-$Sr_2FeMoO_6$ system reveal larger magnetoresistance values in the composite particles than in the individual counterparts [301]. Finally, it has been recently demonstrated that $Fe_3O_4/CoFe_2O_4$ inverse soft/hard core/shell nanoparticles exhibit tunneling magnetoresistance controlled by the magnetic properties of the insulating barrier (i.e., hard $CoFe_2O_4$ shell) rather than by the ones of the conducting counterpart (i.e., soft $Fe_3O_4$ core). Namely, the field at which the magnetoresistance exhibits maxima (usually related to the coercivity of the conductor) is much larger than $H_C$ for $Fe_3O_4$ [254]. This effect is similar to the one observed in bulk samples due to grain boundary effects [462] and allows, to certain extent, to engineer the tunneling magnetoresistance.

Another tentative application for hard-soft core/shell nanoparticles may be in magnetic refrigeration, i.e., magnetocaloric devices based on the change of temperature of a magnetic material upon the application or removal of a magnetic field [463–465]. Given that both soft and hard materials can lead to magnetocaloric effects [463–467] perhaps the combination of both in a core/shell morphology may lead to attractive effects. Bi-magnetic nanocomposites and multilayers (and in particular hard-soft composites) have been shown to result in enhanced magnetocaloric effects [468–471]. Similarly, soft nanoparticles with a large surface anisotropy or spin-glass surface spins have also been proposed as candidates for strong magnetocaloric effects [472,473]. Although no magnetocaloric studies of hard-soft core/shell nanoparticles can be found in the literature, a recent report on $Fe/\gamma\text{-}Fe_2O_3$ core/shell nanoparticles has revealed some interesting features (e.g., entropies changes of opposite sign at different temperatures), with contributions from both the core and the shell [474].

In fact, bi-magnetic core/shell nanoparticles could be used in any application based on magnetic nanoparticles where the coercivity and/or the magnetization need to be fine-tuned. One example could be nanoscale magnetic cellular automata. In lithographed dots magnetic logic using dipolar interactions has already been demonstrated [454,475–477]. It could be envisioned that magnetic logic could be achieved using nanoparticles [478–480], thus considerably reducing the size of the devices. However, this would imply having a $M_S$ sufficiently high to induce large enough dipolar fields to the nearest particles, a low enough $H_C$ to allow the influence of the neighboring particles though dipolar interactions. Nevertheless, the anisotropy of the particles should be sufficiently high to ensure they are magnetically stable at room temperature.. Fulfilling all these requirements may not be trivial for single phase nanoparticles, thus bi-magnetic core/shell nanoparticles could be a suitable choice.

Although additional sophisticated uses cannot be foreseen today, certainly other novel applications will emerge in the future.

## 4. Conclusions

In conclusion, the basic properties and applications of bi-magnetic hard-soft core/shell nanoparticles have been reviewed. The great progress in controlled synthesis and the advances in structural-morphological and magnetic characterization have been highlighted. It has been shown that this type of particles hold a great potential for a range of uses such as permanent magnets, magnetic recording, microwave absorption or biomedical applications. Moreover, other possible applications, like miniaturized magnetotransport devices, have also been proposed. Nevertheless, despite their appealing properties, for certain purposes, e.g., permanent magnets or recording media, there still exist some aspects which need to be addressed before they can be implemented in devices.

**Acknowledgements**

The authors thank A. Quesada, E. Winkler, C. Sangregorio, J.A. de Toro, M.P. Morales and M.D. Baró for the critical reading of the manuscript. This work has been supported by the MAT2010-20616-C02 and MAT2013-48628-R projects of the Spanish Ministerio de Economía y Competitividad (MINECO) and by the 2014-SGR-1015 project of the Generalitat de Catalunya.. M. Estrader wants to thank the Spanish MINECO for a Juan de la Cierva post-doctoral fellowship. A. G. Roca would like to thank Generalitat de Catalunya for financial support under the Beatriu de Pinos fellowship program (2011 BP_B 00209). ICN2 acknowledges support from the Severo Ochoa Program of the Spanish MINECO.